\documentclass[%
 reprint,
 amsmath,amssymb,
 aps, prd,preprintnumbers, superscriptaddress, nofootinbib
]{revtex4-2}

\usepackage{booktabs}
\usepackage{natbib}
\usepackage{graphicx} 
\usepackage{hyperref}
\usepackage{amsmath}
\usepackage{slashed}
\usepackage{xcolor}

\date{\today}

\begin{document}

\preprint{YITP-SB-2026-16}
\preprint{UCI-TR-2026-16}

\title{Auditing anapole dark matter: the limits of model independence in dark matter effective field theory}
\author{Duncan K. Adams}
\email{duncan.adams@stonybrook.edu}
\affiliation{C.N. Yang Institute for Theoretical Physics, Stony Brook University, Stony Brook, NY 11794-3800, USA}
\affiliation{Department of Physics and Astronomy, University of California, Irvine, CA 92697 USA}

\author{Jared Barron}
\email{jared.barron@stonybrook.edu}
\affiliation{C.N. Yang Institute for Theoretical Physics, Stony Brook University, Stony Brook, NY 11794-3800, USA}

\author{Megan H. McDuffie}
\email{megan.mcduffie@stonybrook.edu}
\affiliation{C.N. Yang Institute for Theoretical Physics, Stony Brook University, Stony Brook, NY 11794-3800, USA}
\affiliation{Department of Physics and Astronomy, Stony Brook University, Stony Brook, NY 11794-3800, USA}

\begin{abstract}
Effective field theory is a powerful tool for studying the low-energy behavior of dark sector models without explicit dependence on the high-energy physics. However, bottom-up EFT analyses can miss important processes in UV completions that are relevant for dark matter production and detection. In this work we demonstrate explicitly that for Majorana dark matter coupled to the photon by the dimension-six anapole operator, different UV completions make qualitatively different predictions at high energy colliders, in the early universe, and in direct detection experiments. Additionally, we show that competitive, and in some cases, leading constraints on anapole dark matter can be derived by searching for signatures of additional particles in the UV completion rather than direct probes of the DM itself. We argue that the UV completions studied in this work have categorically different phenomenologies and that `model-independent' EFT constraints are not always applicable.
\end{abstract}

\maketitle

\section{Introduction}

A wealth of observational evidence has established that approximately one-quarter of the Universe's present day energy density is comprised of dark matter (DM) which behaves as a cold, collisionless fluid on length scales $\gtrsim 1~\mathrm{Mpc}$~\cite{Planck:2018vyg,AtacamaCosmologyTelescope:2025blo,Bergstrom:2000pn}.
A large number of particle physics models have been proposed to explain the observed abundance and properties of dark matter, and while they are all required to reproduce the same large scale cosmological behavior of cold dark matter, these models predict a diverse set of observational signatures in both terrestrial experiments and astrophysical environments. This has led to significant efforts to identify and test predictive benchmark models of dark matter with various experimental techniques and analyses of astrophysical data~\cite{Bertone:2004pz,Feng:2010gw,Cirelli:2024ssz,Bullock:2017xww}. Currently, there still remain a large number of viable models, and although there is a flourishing landscape of experiments that will test them, it will still be many years before some of the most promising dark matter candidates can be definitively tested. 

This veritable zoo of viable dark matter models has motivated the pursuit of a `model-independent' framework to reduce the theory space to a minimal set of interactions dark matter can have with the Standard Model (SM) and derive constraints without relying on the specific details of complete particle physics models. 

Specifically, it is common to apply effective field theory (EFT) techniques, modeling the interactions between the dark matter fields and Standard Model particles with non-renormalizable operators which are generated by a potentially large number of distinct UV completions. This is especially attractive for dark sector models which possess new fundamental interactions between dark matter particles and are weakly coupled to the visible sector with so-called `portal' interactions. In scenarios where typical momenta are much less than the mass scales of portal mediators or the dark force carriers, these particles can be integrated out, leading to some set of low-energy operators relevant for astrophysical and terrestrial probes. In direct detection experiments this scale is the typical momentum of halo dark matter, $q\sim 10^{-3}~m_\chi$. These low-energy operators can be studied without reference to the interactions that generated them. Bounds on the Wilson coefficients of these operators can then potentially be re-cast into constraints on a large number of UV completions at once. One class of such operators are electromagnetic form factors for dark matter. The zero-momentum limit of these form factors can be mapped to non-renormalizable operators, and many studies have placed limits on the allowed values of such operators through various terrestrial and astrophysical probes~\cite{Pospelov:2000bq,Banks:2010eh,Fortin:2011hv,Chu:2018qrm,Chu:2019rok,Chang:2019xva,Eberl:2025kfm}.

One issue with this EFT approach to dark matter phenomenology is that predictive dark sector benchmark models typically require mediator masses that are not too much larger than the dark matter mass, meaning that while the `heavy' dark sector particles being integrated out can be ignored for low-energy processes such as elastic scattering at direct detection experiments or stellar cooling, it is not always the case that they can be ignored in the early universe or at high-energy experiments. Working within the EFT can thus lead to incorrect conclusions about the mechanism by which the dark matter achieved its relic abundance and constraints on the dark sector from high-energy experiments. For example, issues with constraints derived from EFTs were investigated for Higgs portal models in~\cite{Shaikh:2026ixs}.

These issues motivate us to investigate the limitations of using EFTs to predict the cosmological history and experimental signatures of dark matter. One barrier to this effort is that the set of distinct UV completions of a given operator can be quite large, and there is no general method for solving the inverse EFT problem by exhaustively constructing every UV completion for a given spectrum of operators.

This paper presents a case study for a specific EFT by exhaustively working through the plausible UV completions of `anapole dark matter', whose low-energy interactions with the SM are dominated by the dimension-six anapole operator. For each UV completion we study the applicability of existing results from the EFT, the production mechanism(s) of the dark matter, and the strongest constraints that can be derived using knowledge of the UV model. 

The anapole operator is the zero-momentum limit of the anapole form factor, one of the few possible electromagnetic form factors for neutral particles ~\cite{Zeldovich1958}. The anapole form factor is of particular theoretical interest because it is the only electromagnetic moment a Majorana fermion can possess, as a consequence of the CPT theorem~\cite{PhysRevD.32.1266,PhysRevLett.62.852,Kayser:1982br, Nowakowski:2004cv}. There is an extensive literature exploring production and experimental constraints on anapole dark matter~\cite{Ho:2012bg,Gao:2013vfa,DelNobile:2014eta,Latimer:2017lwm,Chu:2018qrm,Chu:2019rok,Florez:2019tqr,Arina:2020mxo,Ibarra:2022nzm,Karoly:2023btk,Bose:2024rbp,Ibarra:2024mpq,Ibarra:2026otc}; however, there has not been a systematic study of the additional effects from the new states that are necessarily present to mediate the anapole interaction. 

We find that there is one renormalizable model that generates the anapole moment at tree-level, with all other completions generating it at loop-level. The regime of validity of EFT-derived limits depends strongly on this distinction. For both loop- and tree-level completions, using the full model allows stronger constraints to be set than would be possible using the EFT. Additionally, in some cases, the cosmological production of dark matter proceeds through processes that cannot be described with the anapole operator alone, resulting in very different predictions of the motivated parameter space. Our work serves as a demonstration of the general principle that the implied model-independence promised by the EFT approach comes with severe caveats, particularly for light dark matter. 

The rest of this paper is organized as follows: in Section~\ref{sec:anapoleInt} we define the anapole operator and discuss the EFT in more detail, including the possible DM production mechanisms and constraints from the current literature. In Section~\ref{sec:tree_level}, we detail the tree-level UV completion and discuss the validity of the EFT constraints for this model as well as its production mechanisms and constraints. In \ref{sec:loop_level}, we describe how to generate the anapole from one-loop UV interactions. We again discuss the validity of the EFT constraints, because the relevant energies differ at one-loop. We detail three exemplary models in \ref{subsec:leptophilic}, \ref{subsec:pure_mcps}, and \ref{subsec:effective_mCPs}. Each model has different particle content and interactions, which lead to distinct dark matter production mechanisms and constraints from the additional BSM particles. 
For each model, we elucidate the relevant production mechanism and identify viable parameter space to achieve the relic abundance of dark matter today. Finally, for each model we re-cast constraints from signatures of the additional BSM particles into novel bounds on the anapole moment of dark matter and compare to existing bounds. In Section~\ref{sec:discussion} we summarize the status of anapole dark matter across the space of models and conclude with some general lessons for studying dark matter with EFT. 

\section{Anapole Dark Matter} \label{sec:anapoleInt}
It is convenient to express the interaction vertex of a neutral spin-$\frac{1}{2}$ fermion $\chi$ with the electromagnetic field through its form factors (see e.g. \cite{Nowakowski:2004cv,Chu:2018qrm}),

\begin{equation}
    \langle \chi(p_2, s_2)| J^\mu_{\rm EM}| \chi(p_1, s_1)\rangle = \bar u(p_2, s_2) \Gamma^\mu (q) u(p_1, s_1), 
\end{equation}
where $q=p_{1}-p_{2}$, and 
\begin{align}
    \Gamma^\mu(q) =& (q^2\gamma^\mu -q^\mu \slashed{q})[V(q^2) -A(q^2)\gamma^5] \\ \nonumber
    &+ i\sigma^{\mu \nu}q_\nu [M(q^2)+iD(q^2)\gamma^5].
\end{align}
The form factors encode information about the charge and current distributions of a composite particle, or the effective interactions of a fundamental particle with photons, generated at a scale much higher than $q$. The electromagnetic moments of $\chi$ are the zero-momentum limits of these form factors: $V(0)$ is the charge radius, $M(0)$ is the magnetic dipole moment, $D(0)$ is the electric dipole moment, and $\mathcal{A}\equiv A(0)$ is the anapole moment. 

If the fermion $\chi$ is Dirac, then all four of its form factors can be non-zero. If it is Majorana, CPT symmetry requires that only the anapole form factor can be non-zero~\cite{Zeldovich1958, PhysRevD.32.1266, PhysRevLett.62.852, Kayser:1982br, Nowakowski:2004cv}. For Dirac fermions, the interactions that generate an anapole moment will generically also generate a magnetic dipole moment, which arises at dimension-five. Barring fine tuning, the observational signals from the magnetic dipole moment will dominate over those of the dimension-six anapole. This has been demonstrated explicitly in the context of direct detection and astrophysical data~\cite{Ibarra:2024mpq,Ibarra:2026otc}. Therefore, the anapole moment will only---and indeed, must---dominate the electromagnetic interactions of Majorana fermions. For the remainder of this work we will assume the particle $\chi$ to be a Majorana fermion. 

In the zero momentum limit, the anapole interaction of a Majorana fermion $\chi$ with the electromagnetic field can be conveniently described by a dimension-six operator,
\begin{equation}
    \mathcal{L}_{\rm anapole} = \frac{\mathcal A}{2} \bar\chi \gamma^\mu \gamma^5 \chi \partial^\nu F_{\mu \nu} .
    \label{eqn:Lanapole}
\end{equation}
From the equations of motion, $\partial^\nu F_{\mu \nu}=e J^{\rm EM}_\mu$, so the anapole interaction can be seen as a contact interaction that couples $\chi$ to the electromagnetic current. Notably, it vanishes for on-shell photons in a vacuum.

The non-relativistic limit of the anapole operator can be described using the DM non-relativistic (NR)-EFT to obtain the interactions of the DM with nucleons and electrons, which are relevant for direct detection experiments~\cite{Fan:2010gt,Krnjaic:2024bdd}. Following the commonly used notation of~\cite{Fitzpatrick:2012ix, PhysRevResearch.2.033195} for the NR-EFT operators, the DM-target interaction generated by the anapole is $2eQ_{T}\mathcal{A}\mathcal{O}_{8} - 2e\widetilde{\mu}_{T}\mathcal{A}\mathcal{O}_{9}$, for $T= {e, n, p}$~\cite{PhysRevResearch.2.033195, Gresham:2014vja, Ibarra:2024mpq} (see Appendix~\ref{app:EFT_mapping} for more details). 

Equation~\ref{eqn:Lanapole} describes a parity-violating interaction between the anapole particle $\chi$ and the electromagnetic field~\cite{Zeldovich1958}. The underlying UV theory must therefore involve parity violating interactions between $\chi$ and particles which couple to the photon. In principle, $\chi$ could also be a composite particle possessing an anapole moment. Analogous constructions have been studied for electric or magnetic dipole moments in strongly coupled hidden sectors~\cite{Antipin:2015xia}. For this work, we restrict our attention to perturbatively coupled theories.

To maintain invariance under the full unbroken SM gauge group, the new physics generating the anapole moment should couple to the hyper-charge gauge boson $B^{\mu}$ rather than the photon. Therefore, to obtain the anapole moment of Equation~\ref{eqn:Lanapole}, the photon field strength $F^{\mu\nu}$ in the anapole operator should be replaced with the hyper-charge gauge boson field strength $B^{\mu\nu}$, generating couplings to both the photon and $Z$ boson:

\begin{equation}
\begin{aligned}
    \mathcal{L}_{\rm anapole} &= \frac{\mathcal A}{2\cos(\theta_{w})} \bar\chi \gamma^\mu \gamma^5 \chi \partial^\nu B_{\mu \nu} 
    \\&= \frac{\mathcal A}{2} \bar\chi \gamma^\mu \gamma^5 \chi \partial^\nu A_{\mu \nu} - \frac{\mathcal{A} \tan(\theta_{w})}{2} \bar\chi \gamma^\mu \gamma^5 \chi \partial^\nu Z_{\mu \nu},  
    \label{eqn:LHypanapole}
\end{aligned}
\end{equation}
whre $\theta_{w}$ is the Weinberg angle. If the new physics generating the anapole operator is only charged under $U(1)_{\rm Y}$, coupling to the $Z$ is inevitable. Generating only the photon coupling would require both fine-tuning and new $SU(2)_{\rm L}$ multiplets. We therefore use the operator of Equation~\ref{eqn:LHypanapole}. At scales $\ll m_{Z}$, only the photon interaction is relevant. The primary effects of including the $Z$ coupling are to introduce a resonance in the freeze-out line at $m_{\chi}=m_{Z}/2$ and introduce the possibility of $Z$ decays to $\chi\chi$ pairs, as we will see in the next sections.

\subsection{Production Mechanisms} \label{subsec:EFTproduction}

For large enough values of $\mathcal{A}$, thermal equilibrium between the SM and DM at early times is established through anapole-mediated processes such as $\chi\chi \leftrightarrow \bar{f}f$. A relic population of $\chi$ particles is then generated through standard thermal freeze-out via p-wave annihilation to Standard Model fermions. The thermal freeze-out target for anapole DM was first calculated in~\cite{Ho:2012bg}, and is approximately 
\begin{equation}
    \mathcal{A}_{\rm th} \sim 0.1\ \mathrm{GeV}^{-2}\ \left(\frac{m_{\chi}}{10\ \mathrm{MeV}}\right)^{-1},
\end{equation}
neglecting deviations due to new annihilation channels becoming kinematically accessible for higher $m_{\chi}$. Updated calculations have been performed in~\cite{Chu:2018qrm,Ibarra:2024mpq}, with good agreement. We use the result of \cite{Ibarra:2024mpq} as our freeze-out target, which assumes a coupling to the $B$ boson and includes the resonance at $m_{Z}/2$\footnote{Our normalization of the anapole moment differs from \cite{Ibarra:2024mpq} by a factor of 1/2. As a consequence, their Dirac freeze-out line applies to Majorana $\chi$ in our convention}.  

As an alternative to thermal freeze-out, UV freeze-in (UVFI) can generate the DM relic abundance at much lower $\mathcal{A}$, provided that the new physics scale lies above the reheat temperature $T_{\rm RH}$. Therefore, this mechanism-- the production of dark matter out of equilibrium with the SM via non-renormalizable operators-- introduces a dependence of the dark matter yield on the reheat temperature~\cite{Elahi:2014fsa,Bernal:2019mhf,Bernal:2025fcl}. When $m_\chi \ll T_{\rm RH}$, the anapole moment required to achieve the correct relic abundance scales as $\mathcal{A}_{\rm UVFI} \propto m_{\chi}^{-1/2}T^{-3/2}_{\rm RH}$. If $m_\chi \gg T_{\rm RH}$, then the DM production is Boltzmann suppressed and instead, $\mathcal{A_{\rm UVFI}} \propto\exp(m_\chi/T_{\rm RH})$. The high-mass cut-off of the UVFI line is set by the requirement that the dark sector should not achieve chemical equilibrium with the SM at $T_{\rm RH}$ or later. We calculate\footnote{Assuming instantaneous reheating} $\mathcal{A_{\rm UVFI}}$ following~\cite{Bernal:2025fcl} and show in Figure~\ref{fig:anapole_eft_constraint} the UVFI targets for reheat temperatures of $4~\mathrm{MeV}$ and $100~\mathrm{MeV}$, consistent with the lower bound on the reheating temperature~\cite{Hannestad:2004px}. 

In the following subsection we review the constraints on $\mathcal{A}$ from different experimental and astrophysical probes that only rely on the presence of the effective operator in Equation~\ref{eqn:Lanapole}, and summarize in Figure~\ref{fig:anapole_eft_constraint}. 

\subsection{Constraints}
\label{subsec:EFTConstraints}
Constraints on dark matter coupled to the SM through the anapole operator from a wide range of experiments and observations have been studied in detail previously~\cite{Chu:2018qrm,Chu:2019rok,Arina:2020mxo,Ibarra:2024mpq,Ibarra:2026otc}. In this section, we briefly review constraints on the EFT anapole operator that exist in the literature. Later, we will return to these constraints in the context of different UV models, where we will highlight which constraints remain valid.  

At colliders, pair production of $\bar{\chi}\chi$ pairs accompanied by initial-state radiation (ISR) emission of a photon or jet has long been used as a search channel for dark matter~\cite{Borodatchenkova:2005ct,Goodman:2010yf,Bai:2010hh,Fox:2011fx,Goodman:2010yf,Fox:2011pm}. Searches for missing transverse energy (MET) plus a photon or jet at BaBar, LEP, and the LHC have been applied to the anapole operator. These high-energy collider constraints provide flat upper bounds on $m_{\chi}$ up to $\sim\sqrt{s}$ for each collider. The LHC bound is strongest, restricting $|\mathcal{A}|<3.2\times 10^{-6}\ \textrm{GeV}^{-2}$ up to $m_{\chi}\sim 200$ GeV, with a collider center-of-mass (CoM) energy $\sqrt{s}=13$ TeV~\cite{Gao:2013vfa,CMS:2017zts,Arina:2020mxo}. This constraint covers the thermal freeze-out line up to $m_{\chi}\sim 200$ GeV. The LEP bound, from a mono-photon search at the L3 experiment, is $|\mathcal{A}|<1.5\times 10^{-5}\ \textrm{GeV}^{-2}$ for $m_{\chi}\lesssim 90$ GeV, using data taken at $\sqrt{s}\approx 188.6 -209.2~\mathrm{GeV}$ GeV~\cite{L3:2003yon,Chu:2018qrm}. The BaBar bound, also from a mono-photon search but with $\sqrt{s}=10.35$ GeV, is $|\mathcal{A}| < 2\times 10^{-3}\ \textrm{GeV}^{-2}$ for $m_{\chi}<1$ GeV~\cite{BaBar:2017tiz,Chu:2018qrm}. 

Another strong constraint arises for $\chi$ coupled to hyper-charge with $m_{\chi}<m_{Z}/2$, in which case Z can decay to $\bar{\chi}\chi$. Measurements of the Z invisible decay width from LEP at $\sqrt{s}=m_{Z}$ set the limit $|\mathcal{A}|<3.8\times 10^{-6}\ \textrm{GeV}^{-2}$, surpassing the LEP mono-photon bound in this region~\cite{ALEPH:2005ab,Chu:2018qrm,ParticleDataGroup:2018ovx}.

One final constraint originating from high-energy colliders comes from measurements of the running of the fine structure constant, summarized using the parameter $\Delta r$~\cite{ParticleDataGroup:2018ovx}. Contributions to the photon vacuum polarization at the scale $|q^{2}|\sim m_{Z}^{2}$ lead to the limit $|\mathcal{A}|<3.2\times 10^{-5}\ \textrm{GeV}^{-2}$~\cite{Chu:2018qrm}.

Other terrestrial searches at fixed target experiments such as NA64 at CERN and mQ at SLAC yield sub-dominant limits to those from colliders, limiting $|\mathcal{A}|<0.1\ \textrm{GeV}^{-2}$ for $m_{\chi}\lesssim 1$ GeV~\cite{NA64:2017vtt,Chu:2018qrm}. 

Many direct detection experiments are sensitive to the anapole moment in the GeV--TeV mass range. The direct detection constraints were most recently computed in \cite{Ibarra:2024mpq}, including nuclear recoil experiments, Migdal effect searches, and some electron recoil searches that constrain lower masses. Generally, direct detection searches are most sensitive around $m_{\chi} \approx 10-1000\ \text{GeV}$, reaching down to $|\mathcal{A}|\sim 2\times 10^{-6}\ \text{GeV}^{-2}$. These constraints intersect the thermal freeze-out line from $m_{\chi} \sim 50-150\ \text{GeV}$. 

Astrophysical probes provide a complementary avenue to constrain the anapole moment. High-temperature systems can produce the DM on-shell up to $m_{\chi}\sim T$ which subsequently escape, altering the luminosity or evolution of the system. The most powerful of these systems is the supernova SN1987A~\cite{Hirata:1987hu,Bionta:1987qt,Alekseev:1987ej}. Production of new light particles which escape the supernova without interacting leads to excess cooling, creating a tension with the inferred neutrino luminosity of SN1987A~\cite{Chang:2016ntp, Chang:2018rso}. The core temperature of the supernova is $\sim30~\mathrm{MeV}$, allowing production of particles with masses $m \lesssim 100~\mathrm{MeV}$ when the Boltzmann tail is taken into account. In~\cite{Chu:2019rok}, this constraint was applied to the anapole interaction, setting an upper bound of $|\mathcal{A}|<1.3\times 10^{-7}\ \textrm{GeV}^{-2}$ for $m_{\chi}<0.13$ GeV. For $|\mathcal{A}|\gtrsim9\times 10^{-4}\ \textrm{GeV}^{-2}$ the mean free path of the $\chi$ particles is small enough that they scatter before escaping the supernova remnant so that the cooling argument does not apply, leading to an upper boundary on the region constrained by SN1987A. Limits on energy loss in red giant stars, horizontal branch stars, and the Sun due to production of $\bar{\chi}\chi$ pairs have similarly been used to set bounds for $m_{\chi}\lesssim 10-100$ keV~\cite{Chu:2019rok}. These constraints reach down to $|\mathcal{A}|\sim 6\times 10^{-4}\ \textrm{GeV}^{-2}$, dominated by red giants. Bounds on excess heating in white dwarf stars due to dark matter capture and annihilation constrain $|\mathcal{A}|<6\times 10^{-6}\ \textrm{GeV}^{-2}$ for $m_{\chi}\gtrsim 100$ MeV~\cite{Ibarra:2026otc}. 

\begin{figure}
    \centering
    \includegraphics[width=1.0\linewidth]{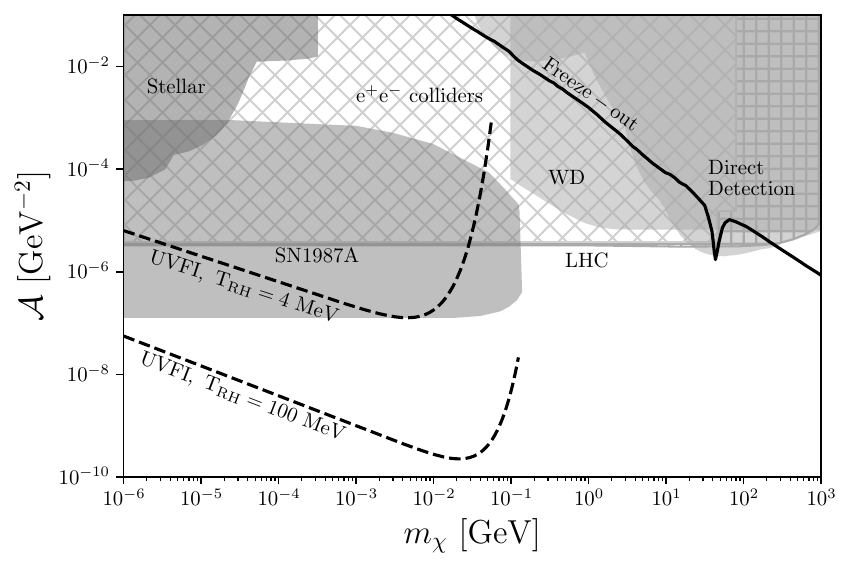}
    \caption{Constraints and relic abundance targets for anapole DM within the EFT framework. Constraints from SN1987A and stellar cooling~\cite{Chu:2019rok}, direct detection~\cite{Ibarra:2024mpq}, and white dwarf heating~\cite{Ibarra:2026otc} are shaded. Leading constraints from LEP~\cite{Chu:2018qrm} and the LHC~\cite{Arina:2020mxo} are hatched. The thermal freeze-out line for anapole DM coupled to hypercharge is shown in solid black, and UV freeze-in lines with $T_{\rm RH}=4,\ 100$ MeV are shown in dashed black.}
    \label{fig:anapole_eft_constraint}
\end{figure}

We display the leading terrestrial and astrophysical constraints on the anapole operator in~Figure~\ref{fig:anapole_eft_constraint} along with the thermal freeze-out line and two example UV freeze-in lines. The thermal freeze-out line appears entirely excluded by collider and direct detection experiments for $m_{\chi}\lesssim 200$ GeV, while UV freeze-in is excluded only by SN1987A, and only for $T_{\rm RH}$ close to the lower bound of $\sim 4$ MeV. Reheat temperatures of $T_{\rm RH}=100$ MeV and above are entirely unconstrained. 

In our exploration of renormalizable models that generate the anapole moment, we will interrogate the validity and applicability of these constraints derived within the EFT, each of which necessarily assumes that the new physics scale is far above the momentum transfer of the process driving the constraint. 

\section{Tree-level completion} \label{sec:tree_level}

There is a unique mechanism that generates the anapole moment at tree-level, which is as follows: the Majorana dark matter $\chi$ has an axial-vector coupling to a massive, kinetically mixed gauge boson $A'$. This $A'$ could be the dark photon of a $U(1)_{\rm D}$ gauge group, or the gauge boson of a gauge extension of the Standard Model, such as gauged baryon minus lepton number $U(1)_{\rm {B-L}}$. The dark sector Lagrangian is,

\begin{equation} \label{eq:treelevelanapole}
\begin{aligned}
    \mathcal{L} \supset & -\frac{1}{4}F'_{\mu\nu}F'^{\mu\nu} -\frac{\epsilon}{2}F_{\mu\nu}F'^{\mu\nu} - \frac{1}{2}M^{2}A'^{2}_{\mu} \\
    & + \frac{1}{2}\bar{\chi}(i\slashed{\partial} - gA'_{\mu}\gamma^{\mu}\gamma^{5} - m_{\chi})\chi,
\end{aligned}
\end{equation}
where $F'_{\mu \nu}$ is the stress-energy tensor of the massive dark photon and $M$ is its mass. 
After an integration by parts, the kinetic mixing term can be re-written $\frac{\epsilon}{2}A'_{\mu}\partial_{\nu}F^{\mu\nu}$, and upon integrating out the dark photon, the DM Lagrangian becomes,
\begin{equation}
    \begin{aligned}
        \frac{1}{2}\bar{\chi}(i\slashed{\partial} -  m_{\chi})\chi&+\frac{1}{2}\frac{\epsilon g}{M^{2}}\bar{\chi}\gamma^{\mu}\gamma^{5}\chi \partial_{\nu}F^{\mu\nu}\\&+\frac{g^{2}}{4M^{2}}(\bar{\chi}\gamma^{\mu}\gamma^{5}\chi)^{2} . 
    \end{aligned}
\end{equation}
We can thus identify the anapole moment as $\mathcal{A} = \frac{\epsilon g}{M^{2}}$. Equivalently, by rotating the kinetic mixing away and not integrating out the dark photon, this model can be viewed as elastic Majorana dark matter~\cite{Berlin:2018bsc}, as recognized in~\cite{Fitzpatrick:2010br,Arias:2019iwo}.

Equation~\ref{eq:treelevelanapole} is not a fully consistent theory on its own, and there are several considerations that require additional model-building, given the ingredients already described. A dark Higgs $\Phi$ with charge $-2Q_{\chi}$ under the $U(1)_{\rm D}$ gauge group is necessary to generate the dark photon mass and Majorana mass while maintaining gauge invariance. Also, at least one additional Weyl fermion charged under the $U(1)_{\rm D}$ is necessary for anomaly cancellation. The dark Higgs Lagrangian is, 
\begin{equation}
    \begin{aligned}
            &|D_{\mu}\Phi|^{2} + \mu^{2}|\Phi|^{2} - \frac{\lambda}{4}|\Phi|^{4} - \frac{\lambda_{\Phi H}}{4}|\Phi|^{2}|H|^{2} 
    \\&- \frac{y}{2}\bar{\chi}(\Phi P_{L} + \Phi^{\dagger}P_{R})\chi,  
    \end{aligned}
\end{equation}
where $D_{\mu}\Phi = (\partial_{\mu} +2igA'_{\mu})\Phi$. The dark Higgs has a vacuum expectation value $f\equiv\sqrt{2\mu^{2}/\lambda}$ that spontaneously breaks the $U(1)_{\rm D}$ gauge symmetry. The dark matter $\chi$ obtains a mass $m_{\chi}=yf/\sqrt{2}$ through its Yukawa coupling to the dark Higgs, and the dark photon obtains a mass $M = \sqrt{4g}f$ through the Higgs mechanism. The quartic mixing term, $\lambda_{\Phi H}|\Phi|^{2}|H|^{2}$, can induce Higgs invisible decays and other deviations in electroweak precision observables~\cite{Robens:2015gla,Ilnicka:2018def,ATLAS:2023tkt}. 

A theory with one Majorana fermion with axial-vector coupling is anomalous. The minimal solution is to add one more Weyl fermion $\xi$ with opposite charge under the $U(1)_{\rm D}$ gauge group. This fermion can be made cosmologically irrelevant by giving it a large Majorana mass through a Yukawa coupling to the dark Higgs. However, a Dirac mass term $m_{\rm D}\bar{\chi}\xi$ mixing $\chi$ and $\xi$ is also allowed by symmetries. If $m_{\rm D}\ll m_{\chi}$, this Dirac mass can be ignored. If not, the DM becomes pseudo-Dirac with inelastic coupling to the $A'$, resulting in qualitatively different phenomenology. 

For $\lambda_{\Phi H}$ and $m_{\rm D}$ sufficiently small, and $m_{\xi}$ sufficiently large, we can avoid complications that would alter the phenomenology of the dark sector. With these model-building considerations in mind, the only particle in the dark sector that is relevant for direct detection is $\chi$, which interacts with the SM via an anapole moment. We now discuss the DM production mechanism and constraints on this model. 

If the dark photon is significantly heavier than the DM, cosmological production of dark matter proceeds through thermal freeze-out or UV freeze-in mediated by the anapole operator, as discussed in Section~\ref{subsec:EFTproduction}. Equivalently, this process can be viewed as annihilation mediated by the heavy dark photon to SM fermions which possess a dark milli-charge $\epsilon g$. For $M<m_{\chi}$, secluded freeze-out through annihilation of $\chi\chi\rightarrow A'A'$ followed by decays of $A'$ to SM particles is a possible channel for freeze-out, though subject to bounds from the CMB on exotic energy injection~\cite{Krnjaic:2025noj}.

With a renormalizable completion for anapole dark matter now specified, it is worth examining the limitations of the EFT framework used to compute the previously discussed predictions and constraints. For a process with momentum transfer $|q^{2}|$, the lowest dark photon mass for which the EFT is conceivably valid is $M^{2}=|q|^{2}$. For thermal freeze-out, the relevant scale is $|q|^{2}\sim m_{\chi}^{2}$. Requiring that the EFT is valid at $\mathcal{A}_{\rm th}$ leads to the upper bound $m_{\chi}\lesssim  (\epsilon g)1\ \rm{TeV}$. 

By setting $\epsilon g$ to the perturbative limit of $\sqrt{4\pi}$, we can conduct the most conservative possible check of EFT validity for each upper bound on $\mathcal{A}$ by requiring that $\mathcal{A}_{\rm max}\equiv \epsilon g/|q|^{2}$ exceeds that upper bound, using the appropriate $|q|^{2}$ for that observable. If it does, then there exists some parameter space in the UV model for which both of the following hold: the EFT can be used to predict that observable, and that parameter space is ruled out. 

The astrophysical bounds from stellar energy loss and SN1987A easily satisfy the EFT validity condition. Stellar systems with $|q|\sim 100$ keV have $\mathcal{A}_{\rm max}^{\rm stellar}\sim 10^{8}\ \textrm{GeV}^{-2}$, and SN1987A with $|q|\sim 100$ MeV has $\mathcal{A}_{\rm max}^{\rm SN1987A}\sim 10^{2}\ \textrm{GeV}^{-2}$, both many orders of magnitude greater than their respective upper bounds on $\mathcal{A}$. For direct detection searches, $|q|\sim v_{\chi}^{2}m_{\chi}\sim 10^{-6}m_{\chi}$ and $\mathcal{A}_{\rm max}^{\rm DD}\gtrsim 10^{6}\ \textrm{GeV}^{-2}$ for $m_{\chi}\lesssim 1$ TeV, so the EFT is valid. Annihilation in white dwarfs occurs at $|q|\sim m_{\chi}$, so $\mathcal{A}_{\rm max}^{\rm WD}\sim \sqrt{4\pi}/m_{\chi}^{2}$. Equating this to the bound from white dwarf heating, we find that the EFT validity condition can only be satisfied for $m_{\chi}\lesssim 1.5$ TeV. 

Unsurprisingly, high-energy colliders are the most likely experiments to encounter issues of EFT validity, as has been studied in the context of DM searches at colliders before~\cite{Fox:2011pm,Busoni:2013lha,Bauer:2016pug}. For s-channel production of $\bar{\chi}\chi$ via the dark photon mediator, the momentum transfer scale is $\sim\sqrt{s}$.

The LHC proton-proton center of mass energy is 13 TeV, but the $\sqrt{s}$ of individual parton scatterings is continuously distributed, controlled by the parton distribution functions. Previous studies have found that the EFT framework is valid for $M\gtrsim 5$ TeV~\cite{Bauer:2016pug} at the LHC, yielding $\mathcal{A}_{\rm max}^{\rm LHC}\sim 10^{-7}\ \textrm{GeV}^{-2}$. This is less than the LHC mono-jet limit of $\mathcal{A}<3.2\times 10^{-6}\ \textrm{GeV}^{-2}$, indicating that there is no perturbative parameter space for which the EFT bound is valid. Instead, the LHC mono-jet constraint should be computed using the full theory, including the propagating $A'$. Because the partonic $\sqrt{s}$ is continuously distributed, $A'$ with $M\ll\sqrt{s_{\rm tot}}$ can be resonantly produced at the LHC, enhancing the production cross-section. 

At LEP, $\sqrt{s}$ is $m_{Z}$ or $\sim 200$ GeV, leading to $\mathcal{A}_{\rm max}^{\rm LEP}\sim 10^{-4}\ \textrm{GeV}^{-2}$, larger than the LEP bounds at $\sim 10^{-5}\ \textrm{GeV}^{-2}$. At BaBar, $\sqrt{s}=10.35$ GeV, so $\mathcal{A}_{\rm max}^{\rm BaBar}\sim 10^{-2}\ \textrm{GeV}^{-2}$, larger than the BaBar bound at $\sim10^{-3}\ \textrm{GeV}^{-2}$. Thus, for both BaBar and LEP there are values of $\epsilon g<\sqrt{4\pi}$ and $M>\sqrt{s}$ corresponding to anapole moments $\mathcal{A}$ that are ruled out. However, these bounds do not apply to lower $M$ and $\epsilon g$ that realize the same anapole moment, and in particular, do not robustly rule out the thermal freeze-out line. Requiring the anapole EFT to both be valid at LEP and realize the thermal freeze-out line requires $\mathcal{A}_{\rm LEP}\geq\mathcal{A}_{\rm th}$, which implies that $m_{\chi} \gtrsim 40(\epsilon g)^{-1}\ \textrm{GeV}$. For $\epsilon g \lesssim 0.3$, this lower bound on $m_{\chi}$ exceeds 100 GeV, and none of the freeze-out line putatively covered by LEP constraints is actually accessible within the EFT. 
The same requirement applied to BaBar yields the lower bound $m_{\chi}\gtrsim 100(\epsilon g)^{-1}$ MeV. For $\epsilon g \lesssim 0.1$, this lower bound exceeds 1 GeV, the limit of sensitivity of the BaBar analysis. 

The dependence of the collider cross-section on the UV model parameters qualitatively changes when the dark gauge coupling and dark photon mass are simultaneously lowered while keeping $\mathcal{A}$ constant, and therefore also the relic abundance, fixed. When the dark photon mediator is light enough to be produced on-shell with $M<\sqrt{s}$, the production rate of dark matter is proportional to $\epsilon^{2}$, not $\mathcal{A}^{2}$. It is then no longer possible to place a unique upper bound on $\mathcal{A}$. These examples illustrate a broader lesson; constraints on an EFT operator obtained under the assumption of EFT validity at some high scale cannot be robustly applied to processes mediated by the same operator at a lower scale.

As we have demonstrated, robustly determining the status of constraints on the freeze-out line requires using the full theory, which in this case is equivalent to the minimal dark photon-mediated Majorana dark matter scenario. Constraints on this model have already been thoroughly studied and reported under the conservative assumption that that $M=3m_{\chi}$~\cite{Berlin:2018bsc,Krnjaic:2025noj}. As long as $M\ll |q|^{2}$, the production cross-section of $A'$ is independent of $m_{A'}$, and larger $M$ requires larger coupling to produce the correct relic abundance, with increased production cross-section. Larger $M/m_{\chi}$ is therefore more tightly constrained in the freeze-out scenario. Constraints on the DM-electron scattering cross-section from~\cite{Krnjaic:2025noj} can be mapped to the anapole moment with the definition, 
\begin{equation}
    \sigma_e = \frac{16 \pi \alpha \alpha_D \epsilon^2 \mu^2_{\chi e}}{(m^2_A + \alpha^2 m^2_e)^2} \rightarrow 16 \alpha \mathcal{A}^2 \mu^2_{\chi e} .
\end{equation}
Some of the strongest bounds come from accelerator experiments, such as NA64, MiniBooNE, BaBar, and LEP, using the full theory~\cite{NA64:2024klw,MiniBooNEDM:2018cxm,Essig:2013vha,Fox:2011fx}.Future results from Belle-II and LDMX are projected to constrain the thermal freeze-out line completely~\cite{Berlin:2018bsc}. Constraints on the anapole operator from direct detection, SN1987A, and stellar cooling also apply to this model directly.

\section{Loop-level completion} \label{sec:loop_level}

To continue exploring UV completions of the anapole DM model, we must go beyond tree-level. At one loop, the minimal realization of this scenario requires the introduction of a boson and fermion which couple to $\chi$ through a parity-violating Yukawa interaction, and are directly or effectively charged under $U(1)_{\rm EM}$. The fermion $\psi$ is spin-1/2, and the boson can be a complex scalar $\phi$ or vector $V$. The vector boson is assumed to acquire its mass through spontaneous symmetry breaking, and is accompanied by a goldstone boson $G$. The Yukawa interactions with $\chi$ are: 
\begin{align} \label{eqn:minimal_anapole}
    \mathcal L_{\mathcal A S} &= \bar\chi[y_LP_L + y_RP_R]\phi^*\psi + \rm h.c. \\ \nonumber 
    \mathcal L_{\mathcal A V} &=  \bar\chi \gamma^\mu [y_LP_L + y_RP_R]\psi V_\mu^+ \\ \nonumber  &~~+\bar\chi \gamma^\mu [y^G_LP_L + y^G_RP_R]\psi G^+ + \rm h.c.
\end{align}
For processes with momentum transfer $q^{2}\ll m_{\psi}^{2},\ m_{\phi}^{2},\ m_{V}^{2}$, these particles can be integrated out, resulting in the operator of Equation~\ref{eqn:Lanapole}~\cite{Ibarra:2022nzm}. The Wilson coefficient in the scalar case is as follows:

\begin{equation} \label{eqn:anapole_moment}
    \mathcal A_S = \frac{-e Q}{96\pi^2 m^2_\chi}\big[|y_L|^2 - |y_R|^2 \big]\mathcal F_S(\frac{m_\psi}{m_\chi}, \frac{m_\phi}{m_\chi}) , 
\end{equation}
where $Q$ is the charge of $\phi$ and $\psi$, and we give the explicit expressions for $\mathcal{F}_{S}$ and the vector-coupled anapole moment in Appendix~\ref{app:anapoleformfactor}. In the limit $m_{\rm heavy}\equiv m_{\psi}=m_{\phi} \gg m_{\chi}$,
\begin{equation}
\mathcal{A}_{S} \approx -\frac{eQ}{192 \pi^{2} m_{\rm heavy}^{2}} \big(|y_L|^2 - |y_R|^2 \big).
\label{eq:anapole_limit}
\end{equation}
In this limit, the vector-coupled anapole moment is $\mathcal{A}_{V}=15\mathcal{A}_{S}$. When $m_{\psi}\neq m_{\phi}$, the anapole moment acquires a logarithmic dependence on $(m_{\psi}/m_{\phi})^{2}$, and scales as $m_{\rm heavy}^{-2}\equiv~\max(m_{\psi},m_{\phi})^{-2}$. Note that a non-zero anapole moment requires that the left- and right-handed Yukawa couplings $y_{L}$ and $y_{R}$ differ in magnitude, reflecting its parity-violating nature. Without loss of generality, we will assume that $y_{L}=0$ and take $y\equiv y_{R}$. 

The suppression of $\mathcal{A}$ by a loop factor $\sim 1/192\pi^{2}$ in the loop-level model exacerbates the issue of EFT invalidity in comparison to the tree-level model, motivating us to re-examine the bounds of Section~\ref{subsec:EFTConstraints}. In the vector-fermion model with $m_{V}=m_{f}$, which yields an anapole moment 15 times larger than the scalar-fermion model, we can derive the largest $m_{\chi}$ for which the anapole EFT is valid with perturbative couplings. This bound is 
\begin{equation}
    m_{\chi} \lesssim (0.05\ \mathrm{GeV})\  Q y^{2} \left(\frac{m_{\rm heavy}/m_{\chi}}{10} \right)^{-2}
\end{equation}
At the perturbative limit $Q=\sqrt{4\pi}/e$, $y=\sqrt{4\pi}$ and with the lightest possible $m_{\rm heavy}=m_{\chi}$, the absolute upper bound is $m_{\chi}\sim700~\mathrm{GeV}$. The resonance at $m_{Z}/2$ provides a narrow band of $m_{\chi}$ for which thermal freeze-out remains possible at higher $m_{\rm heavy}$ and lower coupling.

As in the tree-level completion, astrophysical constraints from SN1987A and stellar cooling, as well as direct detection constraints, do not suffer from EFT validity concerns. However, the white dwarf heating constraint can only be perturbatively EFT-valid up to $m_{\chi}\sim 35$ GeV. At LEP, $\mathcal{A}_{\rm max}$ is $\sim 10^{-6}\ \textrm{GeV}^{-2}$, below both the mono-photon and Z invisible width upper bounds. Similarly, at BaBar $\mathcal{A}_{\rm max}\sim 10^{-4}\ \textrm{GeV}^{-2}$, below the upper bound set by that experiment. Therefore, not only the LHC bounds but also the lepton collider bounds derived using the anapole EFT are invalid. 

\begin{figure}
    \centering
    \includegraphics[width=1.0\linewidth]{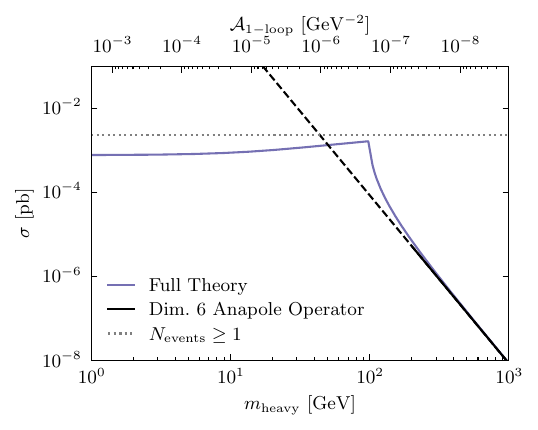}
    \caption{Production cross section of $\chi\chi$ pairs + an ISR photon at LEP for the scalar anapole model in Equation~\ref{eqn:minimal_anapole}, in the kinematic region selected in~\cite{L3:2003yon}, as a function of $m_{\rm heavy}$ at $\sqrt{s} = 200~\mathrm{GeV}$. We take $Q=1$ and $y = \sqrt{4\pi}$. The top axis shows the anapole moment corresponding to $m_{\rm heavy}$ for this choice of parameters. For $m_{\rm heavy} \ll \sqrt{s}$, the cross section computed in the full theory is independent of $m_{\rm heavy}$, whereas the cross section computed using the dimension-six anapole operator rises as $m_{\rm heavy}^{-2}$, leading to a large discrepancy between the number of predicted events between the EFT and the full theory. The cross section computed in the EFT is indicated with a dashed line for $m_{\rm heavy}<200~\mathrm{GeV}$, indicating the region where the EFT approximation is no longer appropriate. The dotted grey line indicates the cross section when the number of expected events exceeds one, assuming a detection efficiency of $70\%$ and a luminosity of $619\ \rm{pb}^{-1}$.  }
    \label{fig:chichigammaxsec}
\end{figure}

To determine if any bound from LEP survives, we explicitly compute the production cross-section of $\chi\chi+\gamma$ at LEP within the loop-level anapole model, taking into account both photon- and Z-mediated diagrams. Additional details of this calculation can be found in Appendix~\ref{app:anapoleformfactor}. We show the cross-section as a function of $m_{\rm heavy}$ in Figure~\ref{fig:chichigammaxsec}. The one-loop cross-section is enhanced compared to the EFT estimate near $m_{\rm heavy}=\sqrt{s}/2$, but becomes independent of $m_{\rm heavy}$ at lower masses. For $\sqrt{s}\gg m_{\textrm{heavy}}$, the total $\chi\chi$ production cross section is,
\begin{equation}
    \sigma_{e^+e^-\rightarrow\chi\chi} \approx \frac{\alpha^2 Q^2 |y|^2}{192\pi^3 s}.
\end{equation}
For $Q=1$ and $y$ at the perturbative limit, an expected eight $\chi \chi$ pairs would be produced at LEP. When accounting for the associated production of a photon in the kinematic range in the event selection used in~\cite{L3:2003yon}, the yield is less than one event. Thus, the LEP search for mono-photon + MET is not sensitive to models that generate the anapole moment at loop-level. 

Without these collider constraints, large regions of anapole parameter space are suddenly unconstrained, including the anapole-mediated freeze-out line below 100 MeV. As we explore specific realizations of dark matter with an anapole moment generated at loop-level, we will find that direct searches for the particles running in the loop lead to even stronger constraints. 

In this section we investigate three different loop-level UV models with particles that interact with $\chi$ as in Equation~\ref{eqn:minimal_anapole} and therefore generate the anapole moment in the low-momentum limit. These models differ in the gauge interactions of the particles $\phi$ (or $V$) and $\psi$, with dramatic consequences for the phenomenology. Extensions to these models are possible but likely only add additional observational signatures. The defining characteristics of the UV models we will discuss in more detail are:
\begin{enumerate}
    \item The fermion $\psi$ belongs to the SM. This restricts the gauge interactions of the scalar $\phi$ depending on whether $\psi$ is a lepton or quark, and whether it is left- or right-handed. 
    \item Both $\phi$ and $\psi$ are new particles. They must be at least as heavy as $\chi$ so that the anapole DM is stable, and they must comply with constraints on new charged states. One way to accomplish this is to assign $\phi$ and $\psi$ a pure milli-charge $Q_\text{mCP}$\footnote{Although we use the terminology `milli-charge' in this work, we will generically consider the range $0 < Q_\text{mCP} < \sqrt{4\pi}$.}.
    \item The mediators $\phi$ and $\psi$ are charged under a new gauge group $U(1)_{\rm D}$, which kinetically mixes with the $U(1)_{\rm Y}$. This results in $\phi$ and $\psi$ that act as effective milli-charged particles. 
\end{enumerate}

Each of these scenarios generates distinctive experimental signatures that result in constraints on the anapole moment when confronted with data. We will also find that the dark matter production mechanism is highly dependent on the details of the UV physics beyond the anapole operator. 

\subsection{Leptophilic $\chi$} \label{subsec:leptophilic}

One possibility is that the fermion $\psi$ in Equation~\ref{eqn:minimal_anapole} is a SM particle. In this case, the scalar $\phi$ must possess appropriate SM gauge quantum numbers so that the Yukawa term is a gauge singlet. Models of this type have been studied in detail as `t-channel dark matter' \cite{Arina:2025zpi} with anapole-mediated interactions with nucleons for Majorana $\chi$~\cite{Garny:2015wea,Ibarra:2022nzm,Ibarra:2024mpq}. We assume here that $\psi$ is a right-handed lepton, $\mu$ or $\tau$\footnote{We do not consider coupling to electrons because the electron is light enough that $|q|^{2}\ll m_{e}^{2}$ does not hold for interactions with nucleons, requiring a momentum-dependent form factor~\cite{Kopp:2014tsa}.}, so that $\phi$ is a singlet under $SU(2)_{\rm L}\times SU(3)_{\rm c}$ with hyper-charge $Y_{\phi}=-Y_{\ell_{\rm R}}$ Additional gauge couplings of $\phi$ only add observable signatures or add interactions that dominate over the anapole moment. For instance, if $\psi$ is a light quark then the DM can scatter with nucleons via the Yukawa coupling at tree-level or loop-level strong interactions to gluons, and the mediator $\phi$ can be produced at tree-level at colliders through the Yukawa coupling. The leptophilic case eliminates these complications and additional signatures, and is sufficient to demonstrate that both the production mechanism and constraints on this model cannot be correctly described within the anapole EFT.  

While $\chi$ has an anapole moment in this model, the relic abundance is controlled by tree-level Yukawa-mediated interactions of $\chi$, not loop-level anapole-mediated annihilation. When the mediator $\phi$ is significantly heavier than $\chi$, t-channel annihilation of $\chi$ to SM fermions dominates thermal freeze-out. When the mass spectrum is compressed with $m_{\phi}/m_{\chi}\lesssim 1.1$, co-annihilations with $\phi$ can significantly affect the thermal relic abundance~\cite{Garny:2015wea}. 
At lower Yukawa couplings $\sim 10^{-6}$, `conversion-driven freeze-out' can control the relic abundance, and for even smaller Yukawas $\sim 10^{-8}$, IR freeze-in can occur~\cite{Garny:2017rxs,Junius:2019dci}.

\begin{figure}
    \centering
    \includegraphics[width=1.0\linewidth]{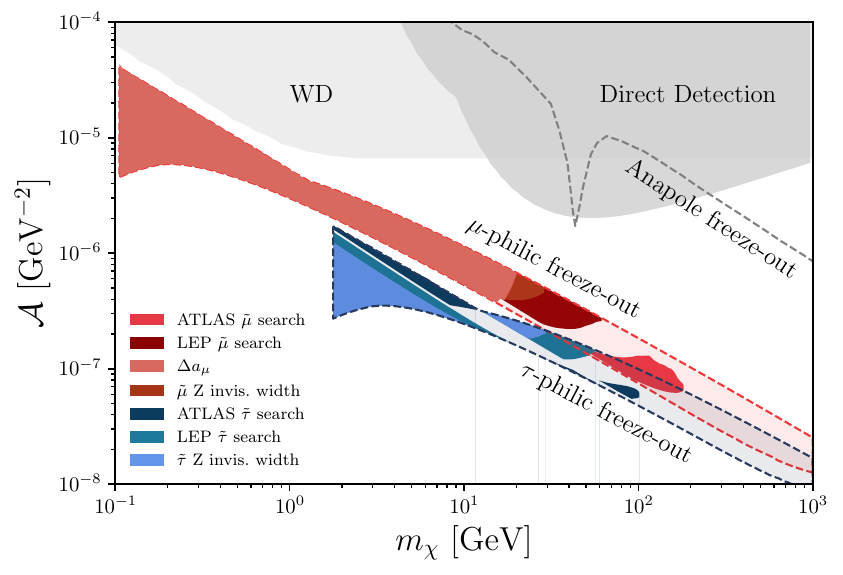}
    \caption{Constraints on anapole DM coupled to SM $\mu$ or $\tau$ leptons. Thermal freeze-out bands for the $\mu$- and $\tau$-philic cases are shown in light red and blue. Constraints on these two models, requiring the correct DM relic abundance, are shaded darker red and blue. These include constraints from SUSY searches, the Z invisible width, and muon $g-2$~\cite{DELPHI:2003uqw,ATLAS:2019lff,ATLAS:2024fub,LEP:2000pgt,Muong-2:2025xyk,Aliberti:2025beg}. Direct detection and white dwarf constraints are shaded light gray, and the anapole-mediated freeze-out line is shown in dashed gray~\cite{Ibarra:2024mpq, Ibarra:2026otc}.}
    \label{fig:anapole_SMcoupled_constraint}
\end{figure}

When cast into the space of $m_{\chi}$ and $\mathcal{A}$, the thermal freeze-out line becomes a band due to the presence of two free parameters in the theory-- the Yukawa coupling $y$ and the scalar mass $m_{\phi}$. The edges of the freeze-out band are set by requiring $m_{\phi}>1.5m_{\chi}$, so that co-annihilations are insignificant, and requiring $y<\sqrt{4\pi}$, so that the theory is perturbative. These bands lie more than an order of magnitude below the prediction from anapole-mediated freeze-out, as shown in Figure~\ref{fig:anapole_SMcoupled_constraint}.  
While direct detection searches for nuclear recoils in the Panda-X and LZ detectors constrain the anapole freeze-out line for $m_{\chi}$ between $\approx 5$ and 100 GeV~\cite{Ibarra:2024mpq}, they are far from probing the correct freeze-out band for these leptophilic models. We find that the thermal relic target outside the co-annihilation regime is completely unconstrained by direct detection. 

Only by considering the UV completion of the anapole EFT are we able to correctly identify the DM production mechanism and make a prediction for the direct detection signature. We now turn to other constraints on these leptophilic models, which already significantly constrain the thermal freeze-out target, unlike direct detection.

Several constraints can be placed on the leptophilic models which are inaccessible from the EFT point of view. The Yukawa term $\bar{\chi}\phi \ell_{R}$ is phenomenologically similar to the coupling of a neutralino, slepton, and lepton in super-symmetric models, so we can reinterpret collider searches for sleptons at both the LHC and LEP to obtain bounds on $m_{\phi}$ and $m_{\chi}$. These searches are sensitive to $m_{\phi}$ up to $\sim 450$ GeV and $m_{\chi}$ up to $\sim 180$ GeV in the $\mu$-philic case, and $m_{\phi}$ up to $\sim 350$ GeV and $m_{\chi}$ up to $\sim 100$ GeV in the $\tau$-philic case~\cite{DELPHI:2003uqw,ATLAS:2019lff,ATLAS:2024fub}. 

Through its hyper-charge, the scalar $\phi$ is coupled to the Z boson. For $m_{\phi}<m_{Z}/2$, $Z\rightarrow \phi^{*} \phi$ decays are kinematically accessible, violating the bound on the Z invisible decay width from LEP~\cite{LEP:2000pgt}. This bound applies to any $m_{\chi}<m_{Z}/2$, so that $\chi$ can be lighter than $\phi$. 

Finally, in the $\mu$-philic case, the new states contribute to the anomalous magnetic moment of the muon. Comparing this contribution~\cite{Garny:2015wea} to the most recent experimental result from the Muon g-2 Experiment~\cite{Muong-2:2025xyk} and SM prediction~\cite{Aliberti:2025beg}, we can robustly exclude $m_{\chi}$ across the thermal freeze-out band up to $m_{\chi}\sim 16$ GeV. 

We recast these constraints into the space of $m_{\chi}$ and $\mathcal{A}$. To see how the constraints apply to the thermal freeze-out band and eliminate the extra degree of freedom, we restrict $y=y_{\rm thermal}$. An additional complication arises because the function $\mathcal{A}(y_{\rm thermal},m_{\chi},m_{\phi})$ is not monotonic in $m_{\phi}$. This can sometimes result in a particular $\mathcal{A}$ being ruled out for one value of $m_{\phi}$, but allowed for another value. We do not constrain such cases, resulting in constraint contours that are disconnected. These constraints are overlaid on the thermal freeze-out bands in Figure~\ref{fig:anapole_SMcoupled_constraint}. 

We find that by leveraging our knowledge of the particles and interactions generating the anapole moment, we can rule out large portions of the true freeze-out band that are far below the sensitivity of any direct detection experiment. The $\mu$-philic case is ruled out entirely up to $m_{\chi}\sim 16~\mathrm{GeV}$, and LHC searches are sensitive to $m_{\chi}\sim180$~GeV. The $\tau$-philic case is almost entirely ruled out up to $m_{\chi}\sim 10$~GeV, and LHC searches are sensitive to $m_{\chi}\sim 100$~GeV.  

From the perspective of the anapole EFT, neither the correct thermal relic target nor the dominant constraints on this model can be computed correctly. The anapole moment is relevant only as the operator generating low-energy interactions with nucleons within the larger, well-studied t-channel DM picture. 

\subsection{Pure milli-charged $\phi$ and $\psi$} \label{subsec:pure_mcps}

A second possibility within the paradigm of loop-level models is that both particles in the anapole-generating loop are BSM particles with electromagnetic charge, which need not be an integer multiple of the SM unit charge $e/3$. In particular, if $Q_{\psi,\phi} \ll 1$, the new particles fall into the commonly studied category of milli-charged particles, permitting their masses to range far below the limits on new states with $Q=1$. There are several mechanisms through which new states can couple to the photon with non-standard charge. We first consider pure milli-charged particles (p-mCPs), which directly couple to SM photons via hypercharge and are singlets under $SU(2)_{\rm L}$. The p-mCP interactions with the photon are
\begin{equation}
\mathcal{L} \supset Q_\text{mCP} e  
\begin{cases}
- A_\mu \bar{\psi}\gamma^\mu \psi & \textrm{(fermion)}\\
- i A^\mu \left( \phi (\partial_\mu \phi^*) - (\partial_\mu \phi) \phi^* \right) \\\hspace{1cm}
        + Q_\text{mCP} e |A_\mu|^2 |\phi|^2  &\textrm{(scalar)} \\
-i V^{\dagger}_{\mu\nu}(A^\mu V^\nu - A^\nu V^\mu) \\\hspace{1cm} + Q_\text{mCP} e A_\nu A_\mu V^{\dagger}_{\nu} V_{\mu} + \textrm{h.c.} & \textrm{(vector)} , 
\end{cases}
\end{equation}
where the p-mCP is $\psi$ for a fermion, $\phi$ for a scalar, or $V$ for a vector. We consider $Q_{\rm mCP}$ ranging from 0 up to the perturbative limit $Q_{\rm mCP}e=\sqrt{4\pi}$, encompassing both the usual milli-charged regime and the possibility that the new states are highly charged. Without loss of generality, for the rest of this section we discuss the scalar-fermion model, although we present results for the vector-fermion model as well.

The existence of pure milli-charged particles requires either that the fundamental unit of charge is smaller than $e$/3, or that the SM electromagnetic gauge group is non-compact so that charge is not quantized. Historical interest in p-mCPs was motivated in part by the possibility of ruling out grand unified theories that predict charge quantization\cite{Georgi:1974sy,Foot:1990uf,Foot:1992ui}. We defer to the next section discussion of effective milli-charged particles (e-mCPs) whose coupling to the photon arises from their interactions with a massless $\rm U(1)_{\rm D}$ gauge boson that kinetically mixes with the SM photon. Such models do not threaten charge quantization. 

In the remainder of this section, we study the phenomenology of DM whose anapole moment is generated by loop-level diagrams involving p-mCPs, focusing on the strong constraints that can be placed on the anapole moment from direct searches for p-mCPs.

Unlike the leptophilic case, dark matter production is indeed mediated by the anapole interaction. The EFT-based predictions of the freeze-out and freeze-in lines in Section~\ref{subsec:EFTproduction} are valid as long as $m_{\phi},m_{\psi}\gg m_{\chi}$ and $m_{\phi},m_{\psi} > T_{\rm RH}$, respectively. 

Many searches have been conducted for mCPs, using cosmological and astrophysical observations as well as terrestrial experiments. For the first time, we re-cast the bounds from these searches into constraints on the anapole moment of $\chi$. We find that even the most conservative mCP constraints set the tightest bounds on the anapole moment for $m_{\chi}\gtrsim 100$ MeV, and are complementary to supernova bounds at lower $m_{\chi}$. 

Searches for mCPs have a long history~\cite{Goldberg:1986nk}. Astrophysical bounds from stellar cooling and SN1987A and accelerator constraints from SLAC were computed and refined in the 1990s~\cite{Golowich:1986tj,Prinz:1998ua,Dobroliubov1990,Mohapatra:1990vq,Bartlett:1990qq,Mohapatra:1991as,Davidson:1991si, Davidson:1993sj,Davidson:2000hf}. More recently, renewed interest in mCPs as a dark matter candidate has motivated a new generation of dedicated mCP experiments, such as JUNO~\cite{ArguellesDelgado:2021lek,Wu:2024iqm}, SUBMET~\cite{Kim:2021eix}, and MilliQan~\cite{milliQan:2021lne, Alcott:2025rxn}, as well as the re-analysis of previous experimental results, for example from LSND~\cite{Magill:2018tbb}, BEBC~\cite{Marocco:2020dqu}, NA64~\cite{Gninenko:2018ter, Gninenko:2025aek}, and SENSEI~\cite{SENSEI:2023gie}. Next-generation experiments such as FORMOSA~\cite{Foroughi-Abari:2020qar,Citron:2025kcy}, LDMX~\cite{Berlin:2018bsc, Gninenko:2025aek}, FERMini~\cite{Kelly:2018brz}, Oscura~\cite{Oscura:2023qch}, and BDX~\cite{Essig:2024dpa, Achenbach:2025ynn} can extend sensitivity to mCPs. 

\begin{figure}
    \centering
    \includegraphics[width=1.0\linewidth]{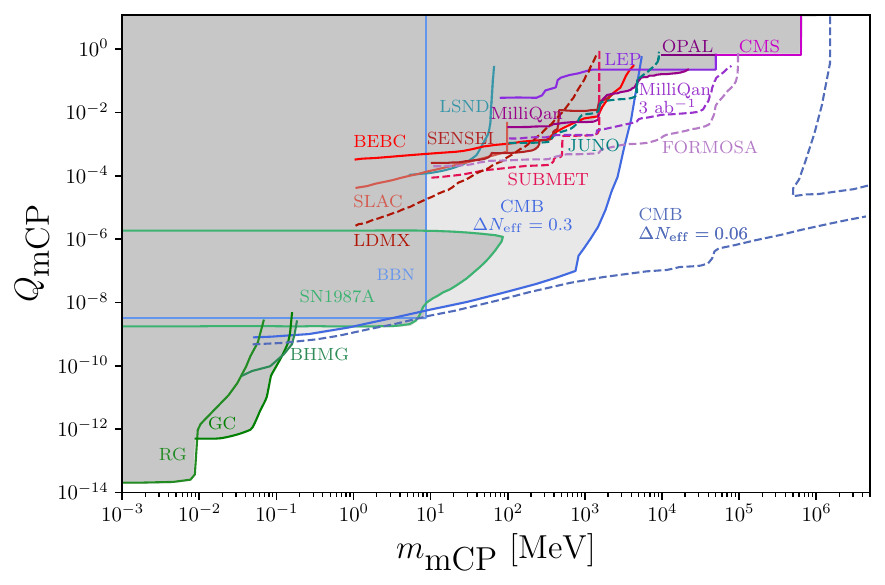}
    \caption{Constraints on milli-charged particles. Astrophysical constraints from Supernova 1987A~\cite{Chang:2018rso}, stellar energy loss in red giants~\cite{Vogel:2013raa}, globular clusters~\cite{Fiorillo:2026psi}, and the black hole mass gap~\cite{Fiorillo:2026vkd} are shown in green. Cosmological constraints from $N_{\rm eff}$ at the time of Big Bang Nucleosynthesis (BBN)~\cite{Boehm:2013jpa,Creque-Sarbinowski:2019mcm} and the CMB~\cite{Vogel:2013raa,Munoz:2018pzp} are shown in blue. CMB projections are shown in dashed blue \cite{Adshead:2022ovo}. Note that constraints from the CMB rely on the existence of a kinetically mixed dark photon, while all other bounds only require the existence of the milli-charged particle. Constraints from fixed target experiments and beam dumps including SLAC~\cite{Prinz:1998ua}, BEBC~\cite{Marocco:2020dqu}, SENSEI~\cite{SENSEI:2023gie}, and projections (in dashed) from LDMX~\cite{Berlin:2018bsc, Gninenko:2025aek} are all shown in red. Constraints from neutrino experiments LSND~\cite{Magill:2018tbb} and JUNO~\cite{ArguellesDelgado:2021lek, Wu:2024iqm} are shown in teal. Collider constraints from LEP~\cite{OPAL:1995uwx,Davidson:2000hf,OPAL:2003zpa,Beacham:2019nyx}, CMS~\cite{CMS:2024eyx}, partial run-3 data from MilliQan~\cite{Alcott:2025rxn}, and projections from FORMOSA~\cite{Foroughi-Abari:2020qar} and MilliQan~\cite{milliQan:2021lne} are shown in purple.}
    \label{fig:mcp_constraint}
\end{figure}

We summarize the world-leading constraints on mCPs in Figure~\ref{fig:mcp_constraint}, as well as projections from currently-running and expected next-generation experiments which will be sensitive to unconstrained parameter space. In this figure, we show constraints on p-mCPs as well as 'effective milli-charged particles' (e-mCPs). All constraints on p-mCPs also constrain e-mCPs; however, e-mCPs are also subject to CMB limits on the presence of additional radiation in the early universe, commonly parameterized as $\Delta N_\textrm{eff}$~\cite{Vogel:2013raa,Munoz:2018pzp}. 

Milli-charge constraints set an upper bound $Q^{\rm bound}_{\rm mCP}(m_{\rm mCP})$, while the anapole moment scales as $\mathcal{A}(m_{\chi}) \sim Q_{\rm mCP}|y|^{2}/m_{\rm mCP}^{2}$ for $m_{\rm mCP}\gg m_{\chi}$. We re-cast the milli-charge constraints in terms of the anapole moment in the maximally conservative way, by setting the Yukawa coupling to $\sqrt{4 \pi}$ and maximizing $\mathcal{A}$ over all allowed $m_\phi, m_\psi$ with $Q_{\rm mCP}$ set to the largest allowed value. We require that $m_{\phi},m_{\psi}>m_{\chi}$ so that $\chi$ is the lightest state in the dark sector. In practice, $\mathcal{A}$ is maximized when $m_{\psi}=m_{\phi}$. Explicitly, the upper bound $\mathcal{A}^{\rm bound}(m_{\chi})$ is defined by 
\begin{equation}
\begin{aligned}
\max_{m_{\psi},m_{\phi}>m_{\chi}}\mathcal{A}\big(&m_{\chi};m_{\phi},m_{\psi},\sqrt{4\pi},\\ &\min(Q^{\rm bound}_{\rm mCP}(m_{\phi}),Q^{\rm bound}_{\rm mCP}(m_{\psi}))\big).
\end{aligned}
\end{equation}

The Yukawa coupling of the dark matter $\chi$ to the mCPs induces a quantum correction to its mass of order $\delta m_{\chi} \sim (y^2/16 \pi^2) m_{\rm mCP}$~\cite{Chang:2019xva}. To avoid fine-tuning of $m_{\chi}$ we can impose the additional condition that $m_{\phi},m_{\psi}\leq 4\pi m_{\chi}$ as a less conservative but more realistic choice. 

\begin{figure*}
    \centering
    \includegraphics[width=0.49\linewidth]{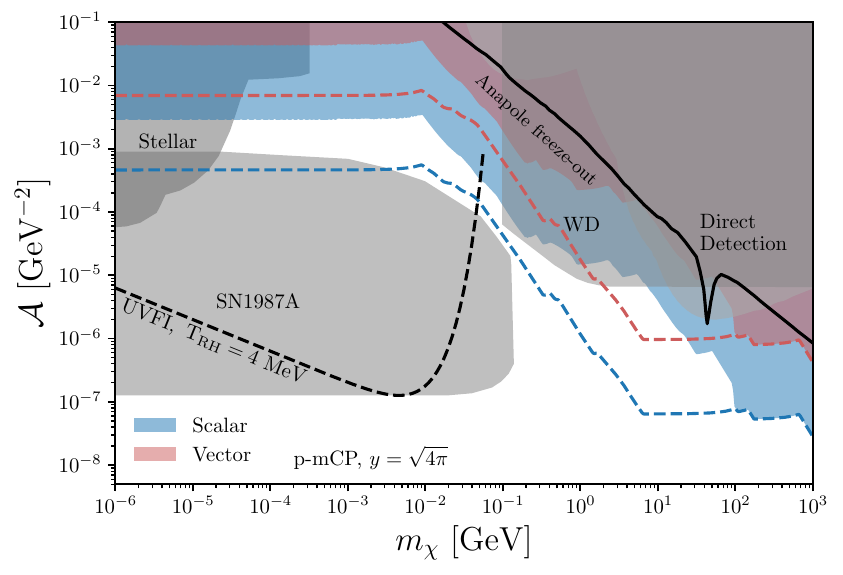}\hfill
    \includegraphics[width=0.49\linewidth]{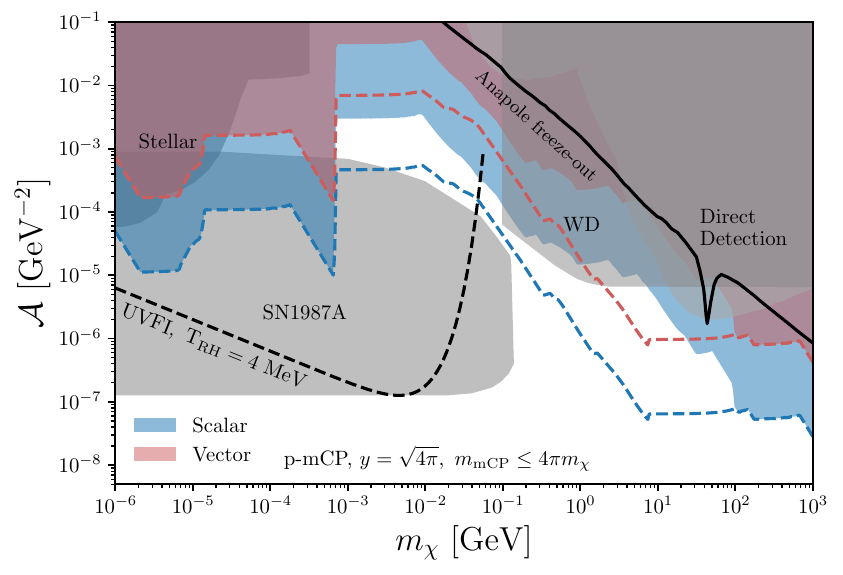}
    \caption{Constraints on the anapole moment from constraints on p-mCPs. We present (\textbf{Left}) the most conservative constraints and (\textbf{Right}) constraints with the naturalness condition $m_{\rm mCP} \leq 4\pi m_{\chi}$. Previous constraints on the anapole operator from SN1987A, stellar cooling, and direct detection are shown in gray~\cite{Chang:2018rso,Chu:2019rok,Ibarra:2024mpq, Ibarra:2026otc}. The anapole-mediated thermal freeze-out line is shown in black and the UVFI line with $T_{\rm RH}=4$ MeV is shown in dashed black. Constraints for a heavy scalar and fermion mCP are shown in blue, and constraints for a heavy vector and fermion mCP are shown in red. Projected constraints for near-future experiments are shown as dashed lines.}
    \label{fig:pmcp_anapole_constraints}
\end{figure*}

We show the re-cast milli-charge constraints on $\mathcal{A}$ both with and without the restriction that $m_{\chi}$ not be fine-tuned in Figure~\ref{fig:pmcp_anapole_constraints}, along with constraints computed in the anapole EFT that are valid for this loop-level completion. Constraints on the vector-fermion model, shown in red, are always a factor of 15 weaker than those for the scalar-fermion model, shown in blue. 

Below $m_{\chi}\sim 100$ MeV the tightest bound on $\mathcal{A}$ is set by SN1987A observations. Milli-charge constraints are complementary to the supernova constraint below 100 MeV, covering most of the parameter space above $\mathcal{A}\sim 10^{-3}\ \textrm{GeV}^{-2}$ where SN1987A data lose sensitivity. Previously, this regime was thought to be ruled out by LEP mono-photon constraints. Below $m_{\chi}\sim 10$ MeV, milli-charge constraints are driven by the intersection of BBN and SLAC constraints. Above 10 MeV, milli-charge constraints come from a variety of fixed target and collider experiments, and above 100 MeV these set the strongest bounds on $\mathcal{A}$. The thermal freeze-out line is ruled out entirely in both the scalar-fermion and vector-fermion model. Above $m_{\chi}\sim 1$ TeV the freeze-out line itself is not perturbatively accessible. UV freeze-in is only constrained by SN1987A data, and only for $T_{\rm RH}\lesssim 10\ \textrm{MeV}$. Near-future experiments targeting milli-charged particles will strengthen bounds on $\mathcal{A}$ by anywhere from one to two orders of magnitude, depending on $m_\chi$. 

The alternative requirement that $m_{\rm mCP}\leq 4\pi m_{\chi}$ impacts p-mCP constraints only for $m_{\chi}\lesssim 1$ MeV, where the most conservative bound on $\mathcal{A}$ comes from milli-charge constraints at $m_{\rm mCP}=10\ \textrm{MeV} > 4\pi m_{\chi}$. The additional requirement of no fine-tuning forces $m_{\rm mCP}$ to lower values, where stronger constraints on $Q_{\rm mCP}$ from stellar evolution apply. Interestingly, next-generation terrestrial p-mCP experiments will not provide better sensitivity below $m_\chi \lesssim 1~\textrm{MeV}$ under the condition of no fine-tuning.

Working with the UV theory including the particles that generate the anapole moment through their interactions with $\chi$ and the photon has afforded us a great deal of insight. We can identify that LEP constraints on the anapole operator, which appear to set the leading constraint between 100 MeV and 10 GeV and rule out the anapole-mediated thermal freeze-out line, in fact do not apply. Instead, even the most conservative constraints from searches for pure milli-charged particles rule out the freeze-out line and set tighter constraints than direct detection experiments.

\subsection{Effectively milli-charged $\phi$ and $\psi$} \label{subsec:effective_mCPs}

Milli-charged particles are perhaps more commonly realized by charging new scalars and fermions under a dark $\rm U(1)_{\rm D}$ gauge group, whose gauge boson (a dark photon) kinetically mixes with the photon. The relevant terms in the Lagrangian are,
\begin{equation}
    \mathcal{L} \supset  - \frac{1}{4}F_{\mu\nu}F^{\mu\nu}-\frac{1}{4}F_{\mu\nu}'F'^{\mu\nu} -\frac{\epsilon}{2}F_{\mu\nu}'F^{\mu\nu},
\end{equation}
where $\epsilon$ is the kinetic mixing parameter between the dark photon (with field strength tensor $F_{\mu\nu}'$) and the SM photon. After a rotation of the photon and dark photon fields to render the kinetic terms diagonal, the $\rm U(1)_{\rm D}$-charged particles gain a coupling to the photon, $\epsilon e_{\rm D}A_{\mu}J^{\mu}_{\rm D}$, where $e_{\rm D}$ is the dark gauge coupling. These particles, which we will call effective milli-charged particles (e-mCPs), have a SM electromagnetic charge $Q_\textrm{mCP} \equiv \epsilon e_{\rm D}/e$. The smallness of the milli-charge arises from the kinetic mixing parameter $\epsilon$, allowing SM charge quantization to be preserved. The anapole moment of $\chi$ is generated in the same way as for p-mCPs, with the simple redefinition of $Q_\text{mCP}$. However, we will find that the existence of the dark photon has remarkable consequences for the possible dark matter production scenarios and constraints on this model. 

Similar to the p-mCP scenario, dark matter with an anapole moment generated by coupling to e-mCPs can be produced through thermal freeze-out as well as UV freeze-in. Since the constraints on e-mCPs are at least as strong as those on p-mCPs, anapole-mediated freeze-out is completely ruled out up to $m_{\chi}\sim$ TeV, beyond which maintaining EFT validity at freeze-out would require non-perturbative couplings. We now present another viable production mechanism that relies on the existence of the dark photon. 

A new dark matter production mechanism for magnetic dipole dark matter was recently introduced in Ref.~\cite{Berlin:2025zut} in which the heavy dark charged particles that generate the magnetic dipole freeze-in and thermalize the dark sector without achieving equilibrium with the Standard Model. The dark charged particles decouple and freeze-out first, leaving the dark sector populated with only the $U(1)_{\rm D}$-neutral dark matter and dark photons. Finally, the dark matter freezes out by annihilating to dark photons. The parameter space that yields the correct dark matter abundance is distinct both from the fully thermalized freeze-out and UV freeze-in scenarios. We adapt this mechanism for e-mCP-mediated anapole DM and identify new theory targets that can realize the correct dark matter abundance today while complying with all constraints. 

Within this scenario, the dark matter $\chi$ can decouple as a hot or cold relic, depending on whether $\langle \sigma v\rangle '_{\chi\chi\rightarrow A' A'}$ is smaller or greater than the Hubble rate at $T' \sim m_{\chi}$. If the dark matter decouples while relativistic, its relic abundance depends on $Q_{\rm mCP}$ and $m_{\rm mCP}$ but not the Yukawa coupling $y$, because the freeze-in of the milli-charged particles controls the energy density of the dark sector. If the dark matter decouples non-relativistically by annihilating to dark photons at loop level, the relic abundance is dependent on the Yukawa coupling, the dark gauge coupling, and both $m_{\rm mCP}$ and $m_{\chi}$, as these all enter the annihilation cross-section. 

The primary difference between the anapole and magnetic dipole is that the anapole operator is dimension-six while the magnetic dipole operator is dimension-five. We can therefore expect additional powers of $1/m_{\rm mCP}$ to appear when adapting the calculations in~\cite{Berlin:2025zut}. The most prominent example is that the thermally averaged cross-section for $\chi\chi\rightarrow A' A'$, in the limit that $m_{\rm mCP} \equiv m_{S} = m_{f} \gg m_{\chi}$, is~\cite{Latimer:2017lwm}

\begin{equation}
    \langle \sigma v\rangle _{\chi\chi\rightarrow A'A'} \approx \frac{e_{D}^{4}y^{4}}{36864 \pi^{5}}\frac{m_{\chi}^{6}}{m_{\rm mCP}^{8}} .
\end{equation}
This cross-section has two more powers of $m_{\chi}/m_{\rm mCP}$ than the analogous cross-section for magnetic dipole dark matter. We find that for anapole DM, the hot relic line which defines the lower boundary of the parameter space that can realize the correct dark matter abundance is 
\begin{equation}
    \mathcal{A}_{\rm HR} \approx \frac{4\times 10^{-9}}{\text{ GeV}^{2}}  |y|^{2}\left( \frac{\text{MeV}}{m_{\chi}}\right)^{2}\left( \frac{m_{\rm mCP}}{m_{\chi}}\right)^{-3/2}.
\end{equation}

Larger anapole moments can also realize the correct relic abundance if the dark matter and dark photon stay coupled until $T'\sim m_{\chi}$. In such cases, if $e_{\rm D}$ is large enough, annihilation of dark matter to dark photons mediates the freeze-out. Within this scenario $\mathcal{A}$ still depends on $\epsilon$, but the relic abundance does not, so there is no unique $\mathcal{A}$ that yields the correct relic abundance. Instead, there is a band of allowed parameter space for freeze-out that is cut off by upper bounds on $\mathcal{A}$ from experiment, to be discussed in the next subsection. At fixed $m_{\rm mCP}/m_{\chi}$, because larger $m_{\chi}$ requires larger $e_{\rm D}$ to freeze-out to the correct abundance, there is an upper bound from perturbativity on $m_{\chi}$ in the cold relic scenario,  

\begin{equation}
    m_{\chi} \lessapprox 50 \text{ GeV}\ |y|^{2} \left(\frac{m_{\chi}}{m_{\rm mCP}}\right)^{4} \sqrt{g^{*}_{s}(m_{\chi}/10)}.
\end{equation}
Thus, at fixed Yukawa coupling and mCP-to-DM mass ratio, this production mechanism of dark sector IR freeze-in followed by freeze-out predicts a bounded region of parameter space that can achieve the dark matter density we observe today.

\begin{figure*}
    \centering
    \includegraphics[width=0.49\linewidth]{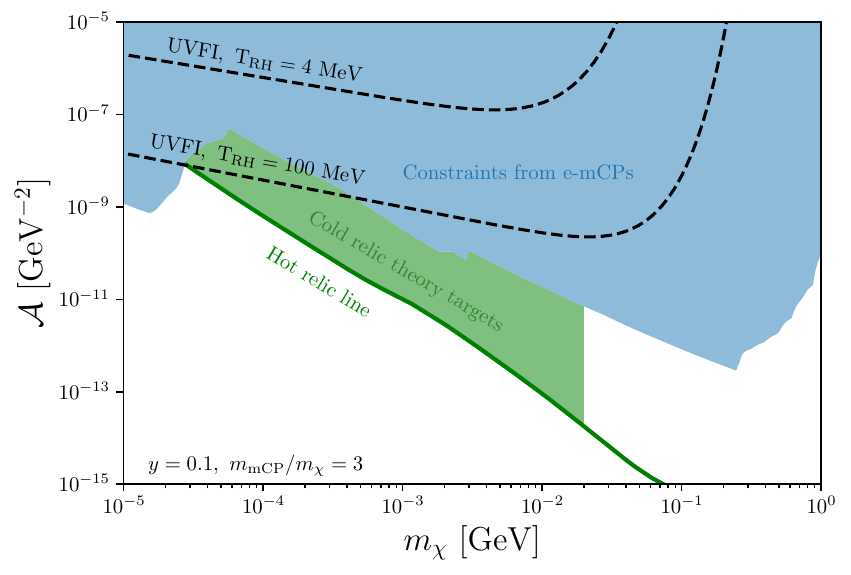}\hfill
    \includegraphics[width=0.49\linewidth]{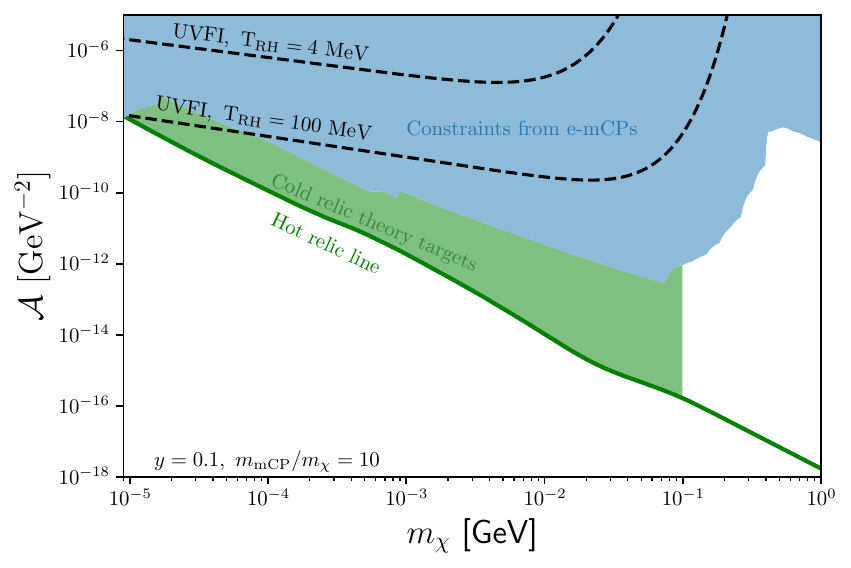}
    \caption{Theory targets and constraints for anapole DM with e-mCPs and a kinetically mixed dark photon in the hidden sector IR freeze-in scenario, with Yukawa coupling $y=0.1$ and \textbf{Left}: mass ratio $m_{\rm mCP}/m_{\chi}=3$ or \textbf{Right}: $m_{\rm mCP}/m_{\chi}=10$. Parameter space where hidden sector freeze-in can achieve the DM relic abundance is shown in green. Constraints from e-mCPs are shown in blue. The UV freeze-in lines for $T_{\rm RH}=4$ MeV and $T_{\rm RH}=100$ MeV are shown in dashed black.}
    \label{fig:hiddensectorFI}
\end{figure*}

We show two examples at different $m_{\rm mCP}/m_{\chi}$ of the target parameter space for this production mechanism in Figure~\ref{fig:hiddensectorFI} along with current constraints from the existence of the e-mCPs. For $y=0.1$, $m_{\rm mCP}/m_{\chi}=3\ (10)$, the cold relic band terminates at $m_{\chi} \lessapprox 20\ (100) \text{ MeV}$. The explicit dependence of the hot relic line and cold relic mass bound on $y$ and $m_{\rm mCP}/m_{\chi}$ necessitate fixing these parameters to uniquely plot the freeze-in target parameter space. 
There are viable anapole DM theory targets in the sub-GeV regime just below the reach of current e-mCP constraints. Tighter bounds on $\Delta N_{\rm eff}$ can only marginally improve constraints in this regime, because $\Delta N_{\rm eff}\lesssim 1$ scales as $\sim Q_{\rm mCP}^{2}$. Nonetheless, the mechanism and theory targets of this IR freeze-in scenario are completely distinct from those of anapole-mediated freeze-out and UV freeze-in, providing another demonstration that knowledge of the full dark sector particle content is necessary to correctly identify the diversity of possible DM production mechanisms and targets in parameter space. 

\begin{figure*}
    \centering
    \includegraphics[width=0.49\linewidth]{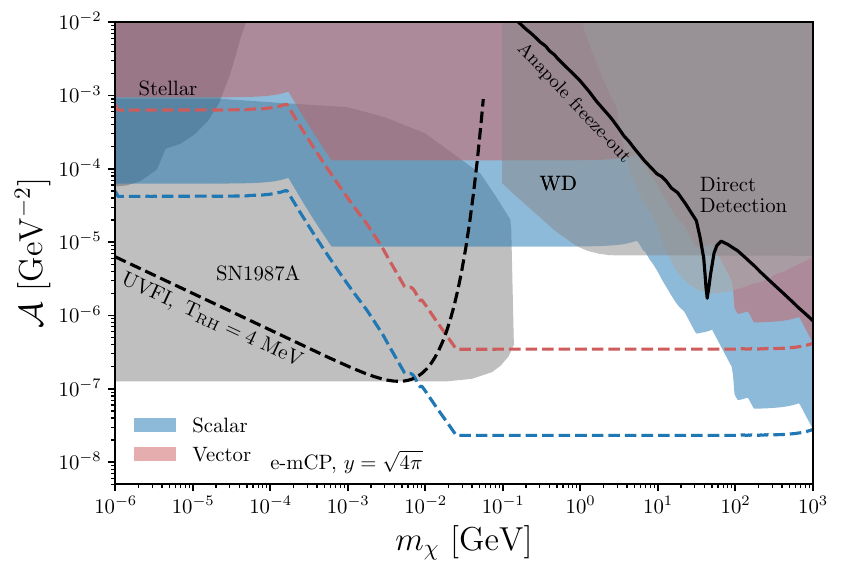}
    \hfill
    \includegraphics[width=0.49\linewidth]{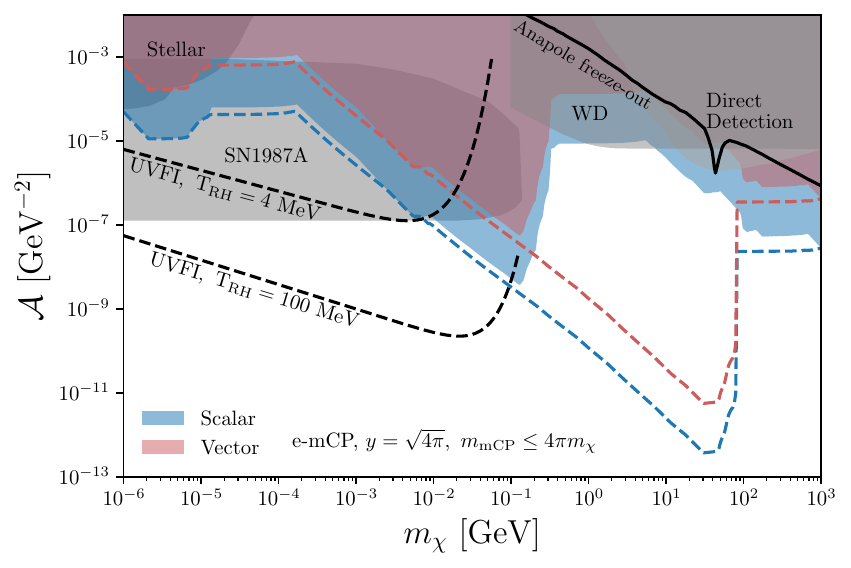}
    \caption{Constraints on the anapole moment from constraints on e-mCPs. We present (\textbf{Left}) the most conservative constraints and (\textbf{Right}) constraints with the naturalness condition $m_{\rm mCP} \leq 4\pi m_{\chi}$. Previous constraints on the anapole operator from SN1987A, stellar cooling, and direct detection are shown in gray~\cite{Chang:2018rso,Chu:2019rok,Ibarra:2024mpq, Ibarra:2026otc}. The anapole-mediated thermal freeze-out line is shown in black and the UVFI line with $T_{\rm RH}=4$ MeV is shown in dashed black. Constraints for a heavy scalar and fermion mCP are shown in blue, and constraints for a heavy vector and fermion mCP are shown in red. Projected constraints for near-future experiments are shown as dashed lines.
    }
    \label{fig:emcp_anapole_constraint}
\end{figure*}

The e-mCPs that mediate the loop share constraints from terrestrial experiments and astrophysical observations with p-mCPs. Additonal constraints apply due to the existence of dark photons, which can contribute non-zero $\Delta N_\text{eff}$ during the CMB epoch~\cite{Vogel:2013raa,Munoz:2018pzp,Adshead:2022ovo}. We also note that the SN1987A bound may change if e-mCPs have self interactions via the dark photon,as shown in~\cite{Fiorillo:2024}. 

The milli-charge portal leads to an irreducible freeze-in population of the e-mCPS which can self-thermalize and generate a population of dark photons, as described previously. The constraint on e-mCPs from the requirement that $\Delta N_{\rm eff}<0.3$, in accordance with Planck 2018 constraints~\cite{Planck:2018vyg}, dominates for $m_{\rm mCP}$ between 10 MeV and 5 GeV. The projected constraint for an S4-like measurement of $\Delta N_{\rm eff}<0.06$ reaches out to $m_{\rm mCP}\sim 1$ TeV, and would improve on current e-mCP constraints by up to seven orders of magnitude in that range. These constraints are shown in Figure~\ref{fig:emcp_anapole_constraint}. 

The current CMB constraint improves on the upper bound on $\mathcal{A}$ compared to p-mCPs by roughly two orders of magnitude for DM masses $\lesssim 10$ GeV. At higher $m_{\chi}$, the constraints are the same. Next-generation measurements of $\Delta N_{\rm eff}$ have the potential to push these bounds almost three orders of magnitude lower for $m_{\chi}\gtrsim 200$ MeV, ruling out $\mathcal{A}$ as low as $10^{-8}\ \textrm{GeV}^{2}$. 

Following the analysis performed for p-mCPs, we also plot constraints under the restriction $m_{\rm mCP}<4\pi m_{\chi}$, corresponding to the `no fine-tuning' limit of the theory. In the range $1~\text{MeV} < m_\chi < 300~\text{MeV}$, the anapole constraint dramatically improves, and the projected constraint from an S4 measurement of $\Delta N_{\rm eff}$ extends as far as $\mathcal{A}\sim10^{-13}\ \textrm{GeV}^{-2}$ for $m_{\chi}\sim 30$ GeV. The Boltzmann tails of UV freeze-in lines at higher $T_{\rm RH}$ may be constrained by future measurements of $\Delta N_{\rm eff}$. 

\section{Discussion} \label{sec:discussion}

In this work we have studied the phenomenology of Majorana fermion dark matter whose low-energy coupling to the Standard Model is dominated by the non-renormalizable anapole operator.

Working initially in the EFT, we have reviewed the thermal freeze-out production mechanism and thermal relic target for this operator. We have also calculated the UV freeze-in yield as a function of the dark matter mass, anapole moment, and reheat temperature, making predictions for the UV freeze-in target at different values of $T_\textrm{RH}$. We have collected the leading constraints on the anapole operator for $m_{\chi}$ ranging from 1 keV to 1 TeV from colliders, direct detection experiments, and astrophysical probes. We find that anapole-mediated UV freeze-in is ruled out by constraints from SN1987A for $T_{\rm RH}\sim 4$ MeV. 

To test the limits of UV model-independence and the utility of the anapole EFT, we have introduced and investigated four classes of dark matter theories whose low-energy scattering with the SM is anapole-dominated. The ingredients of each of these models are chosen to minimally realize the anapole moment while remaining gauge-invariant and anomaly-free. Together, these models span the set of plausible mechanisms for generating the anapole operator. In brief, there is one tree-level completion mediated by a massive, kinetically mixed dark photons, and three loop-level completions. In the first loop-level model, the dark matter possesses a parity-violating coupling to a new scalar and a right-handed SM fermion. In the second, the dark matter couples to two new, pure milli-charged particles. In the third, the dark matter couples to two new particles which couple to a massless, kinetically mixed dark photon. 

For each model we determined the validity of constraints derived in the EFT, identified viable dark matter production mechanisms and target parameter space, and obtained constraints on the anapole moment from searches exploiting signatures of the heavy new physics. We found that leading constraints on the anapole moment from high-energy colliders are not valid for loop-level completions, and cannot robustly rule out the thermal relic target for the tree-level completion. In the cases that dark matter couples directly to a SM fermion or to effective milli-charged particles, we identified alternative dark matter production mechanisms that are not mediated by the anapole operator, with relic targets at significantly lower $\mathcal{A}$ than the anapole thermal relic target. Finally, we found that even the most conservative constraints on the anapole operator from searches for the new physics in each specific model often dominate over constraints derived in the EFT.  In particular, direct detection constraints on the anapole moment are subdominant to other constraints in almost all cases, and for dark matter masses below $\sim 1$ GeV, do not probe realistic dark matter models. For the models that generate the anapole moment at loop-level, experimental searches exploiting the unique signatures of the new physics mediating the anapole interaction are necessary to rule out target parameter space that can realize the dark matter abundance today. We find that the thermal freeze-out line is entirely ruled out if the anapole moment is generated by coupling to milli-charged particles. As a by-product of our analysis, we compile up-to-date and projected constraints on both pure and effective milli-charged particles. 

Our analysis of anapole dark matter exemplifies many of the pitfalls that may restrict the utility of effective field theory techniques in the search for the particle identity of dark matter. First, as a general point, bounds derived within an EFT assuming validity for a process at a high scale cannot be robustly applied to processes in the same EFT at a lower scale. A manifestation of this principle for anapole dark matter is that constraints on $\mathcal{A}$ from LEP searches do not apply if the new particles generating the anapole moment are much lighter than the collider center of mass energy. Furthermore, the EFT parameter space reportedly constrained by high-energy colliders is often unreachable while maintaining EFT validity and perturbative couplings, particularly for loop-level completions. One of the purported benefits of the EFT approach is the ability to constrain many models at once without needing to perform many independent calculations of their experimental signatures. In contrast, we find that there is essentially one perturbative model that generates the anapole moment at tree-level, which is already well-studied as dark-photon mediated Majorana dark matter. Integrating out this dark photon and calculating within the EFT offers no advantage in making predictions or computing constraints on this model. In addition, the theory targets in several of the loop-level completions we consider are dramatically different from the standard anapole thermal freeze-out target. The results of our work demonstrate that it is prudent to carefully examine the assumptions inherent in the use of effective field theory for dark matter, and motivate serious exploration of the landscape of realistic UV completions. Our comprehensive investigation of anapole dark matter serves as an example for applying this methodology to other dark matter models. 

\section{Acknowledgments}
The authors would like to thank Rouven Essig, Patrick Meade, Tanner Trickle, Megan Hott, Merlin Reichard, Majed Khalaf, and Matthew Forslund for useful discussions during the completion of this work. The authors would additionally like to thank the organizers of the MIAPbP workshop:``Fill the Gap: Signatures of Electronic Excitations in Dark Matter and Neutrino Direct Detection Experiments" for their hospitality while the early stages of this project were initialized.
DKA was supported by NSF Grant 2514888, DOE Grant DE-SC0025309, and the Simons Investigator in Physics award MPS-SIP-00010469.
MHM was supported by DOE Grant DE-SC0025309 and the Simons Investigator in Physics award MPS-SIP-00010469. JB was supported by NSF grants PHY-2210533 and PHY-2513893. JB and MHM used the large language models Claude and ChatGPT for some plotting scripts and to verify calculations.

\onecolumngrid

\appendix 

\section{Mapping to non-relativistic EFT} \label{app:EFT_mapping}

For the purposes of predicting dark matter scattering rates in direct detection experiments, it is common to describe DM scattering with the SM through a non-relativistic EFT. This allows one to be agnostic to the UV model, and draw general constraints on a variety of EFT operators. We show here the mapping from the anapole moment to the basis of EFT operators commonly used in the direct detection literature for DM-electron and DM-nucleon scattering experiments. 

Starting with our anapole Lagrangian from Equation~\ref{eqn:Lanapole}, one can replace $\partial^\nu F_{\mu \nu}= e J^{\rm EM}_\mu = e \sum_f Q_f \bar{f} \gamma_\mu f$, where $f$ is a charged SM fermion. We can write down the generic scattering amplitude from the anapole Lagrangian above,
\begin{equation} \label{eqn:scattering_matrix}
\mathcal{M}_\text{anapole} = \mathcal{A} e \left[ \bar{u}_\chi (p')\gamma^\mu \gamma^5 u_\chi(p) \right] J^{EM}_\mu
\end{equation}
The factor of $1/2$ in our definition of the anapole moment is canceled by the additional diagram for self-conjugate $\chi$. We define the momentum transfer $q^\mu = p^\mu - p'^\mu = k'^\mu - k^\mu $ \footnote{The choice to set $\boldsymbol{q} = \boldsymbol{p'} - \boldsymbol{p}$ will change the relative sign of $\mathcal{O}_{9}$.}, with $p (p')$ is the initial (final) DM momentum and $k (k')$ is the initial (final) target momentum. For a given target ($T= e,n,p$), we define the electromagentic current as $J^{EM, T}_\mu$ and use the Gordon identity, to write 
\begin{equation} \label{eqn:gordon_identity}
J^{EM, T}_\mu = \bar{u}_T (k')\left[ F^T_1(q^2) \frac{(k' + k)^\mu}{2m_T} + \left( F^T_1(q^2) + F^T_2(q^2) \right) \frac{i \sigma^{\mu \nu}q_\nu}{2m_T} \right] u_T(k). 
\end{equation}
We can write the form factors in the low momentum limit as $F^T_1(0) = Q_T$, the charge and $F_{1}^{T}(0) + F^{T}_{2} (0) = \widetilde{\mu}_T$, the dimensionless magnetic moment. We can reduce $J^{EM, T}_0 = 2 m_T Q_T$ and $\boldsymbol{J}^{EM,T} = 2 m_T Q_T v_T - 2i \widetilde{\mu}_T \boldsymbol{q} \times \boldsymbol{S}_T$, where $\boldsymbol{v}_T = (\boldsymbol{k} + \boldsymbol{k'}) / 2m_T$, 
and we can also write the DM bilinear with $\bar{u}_\chi (p')\gamma^0 \gamma^5 u_\chi(p) = 2 \boldsymbol{S}_\chi \cdot (\boldsymbol{p}+ \boldsymbol{p'}) = 4 m_\chi \boldsymbol{S}_\chi \cdot \boldsymbol{v}_\chi$, where $\boldsymbol{v}_\chi = (\boldsymbol{p} + \boldsymbol{p'}) / 2m_\chi$ as well as $\bar{u}_\chi (p') \boldsymbol{\gamma} \gamma^5 u_\chi(p) = 4 m_\chi \boldsymbol{S}_\chi$. 
To construct the EFT operators, we must write the bilinear in terms of the Galilean invariant transverse velocity $\boldsymbol{v}^\perp = \boldsymbol{v}_\chi - \boldsymbol{v}_T - \boldsymbol{q}/2 \mu_{\chi T}$, which satisfies $\boldsymbol{v}^\perp \cdot \boldsymbol{q}=0$. Rewriting the anapole amplitude we arrive at, 
\begin{equation}
\mathcal{M}_\textrm{anapole} = 8 e \mathcal{A} m_\chi m_T \left[ Q_T \boldsymbol{S}_\chi \cdot \boldsymbol{v}_\perp - \widetilde{\mu}_T i  \boldsymbol{S}_\chi \cdot \left(\boldsymbol{S}_T \times \frac{\boldsymbol{q}}{m_T} \right) \right] = 4 m_\chi m_T \mathcal{M}_{NR}. 
\end{equation}
We can trivially write the non-relativistic scattering amplitude, relevant for direct detection as, 
\begin{equation}  
\mathcal{M}_\textrm{NR} = 2 e \mathcal{A} (Q_T \mathcal{O}_8 - \widetilde{\mu}_T \mathcal{O}_9) = \sum^{8,9}_{i} c_{i,T} \mathcal{O}_i,
\end{equation}
where $Q_e = -1$, $Q_n = 0$, and $Q_p = 1$, and $\mu_e = -1$, $\mu_n = -1.9$,  $\mu_p = 2.8$, and we define the following two standard EFT operators, 
\begin{align}
    \mathcal{O}_8 &= \boldsymbol{}{S}_\chi\cdot \boldsymbol{v}^\perp \\ \nonumber
    \mathcal{O}_9 &= i \boldsymbol{S}_\chi \cdot \big(\boldsymbol{S}_T \times \frac{\boldsymbol{q}}{m_T} \big).
\end{align}
Therefore, we can construct the mapping from the anapole Lagrangian to the non-relativistic EFT operators with the coefficients $c_{8,e} = - c_{9,e} = - 2 e \mathcal{A}$ for DM-electron scattering~\cite{Catena:2021qsr}, $c_{8,n} = 0, c_{9,n} = - 2 e \widetilde{\mu}_n \mathcal{A}$ for DM-neutron scattering, and $c_{8,p} = 2 e \mathcal{A}, c_{9,p} = - 2 e \widetilde{\mu}_p \mathcal{A}$ for DM-proton scattering~\cite{Fitzpatrick:2012ix, Gresham:2014vja,Ibarra:2024mpq}. 

There are several considerations which strongly impact the sensitivity of direct detection experiments to these operators. For direct detection experiments searching for anapole DM via DM-nucleon scattering, the detector material must be sensitive to spin-dependent interactions. This typically requires nuclear targets with an unpaired nucleon. Isotopes of this kind often have subdominant abundance in detector materials. In addition, to estimate event rates, the nuclear response function of the detector material must be well known~\cite{Fitzpatrick:2012ix, Catena:2015uha}. Isotopes such as xenon-129, xenon-131, germanium-73, and silicon-29 have unpaired neutrons, and so are sensitive to DM-neutron scattering. Because $c_{8,n}=0$, constraints on $\mathcal{O}_{9}$ must be computed to constrain the anapole operator with such experiments in the DM-nucleon scattering channel. This highlights the importance of identifying the NR-EFT operators generated by the relativistic theory. As semi-conductors, germanium and silicon are sensitive to DM-nucleus scattering via the Migdal effect~\cite{Berghaus:2026kmj}. Nuclear response functions for silicon-29 are not well known, inhibiting the current capacity of silicon-based experiments to constrain these operators. Experiments can also have sensitivity to DM-proton scattering. Some notable materials used in current direct detection experiments are fluorine-19, sodium-23,  iodine-127, and aluminum-27. These operators can also be probed through experiments that search for DM-electron scattering. Electron responses have been calculated in many materials to constrain DM-electron NR-EFT interactions, such as in atoms~\cite{PhysRevResearch.2.033195}, crystals~\cite{ Catena:2021qsr} and other exotic materials~\cite{Hochberg:2025rjs}.

We highlight that, generically, DM-electron \emph{and} DM-nucleus direct detection experiments are sensitive to anapole DM. As we have detailed, the anapole Lagrangian maps to a specific pair of coefficients relevant for the non-relativistic EFT. In most direct detection literature, constraints are presented on the coefficients (or equivalently, reference cross section) of a specific operator, assuming all other Wilson coefficients are zero. Generically, this assumption would not be consistent with an anapole motivated UV model. Direct detection constraints on the anapole moment from XENON1T, XENON10, and DarkSide-50 DM-electron scattering searches~\cite{PhysRevResearch.2.033195} highlight the importance of considering both $\mathcal{O}_8$ and $\mathcal{O}_9$ contributions to the scattering rate. Moreover, in DM-nucleus scattering, the direct detection rate includes an additional cross term when both $\mathcal{O}_8$ and $\mathcal{O}_9$ operators contribute~\cite{Fitzpatrick:2012ix}, which introduces a new material response function that can not be neglected when both operators have non-zero coefficients. The anapole operator is the only commonly considered relativistic model that gives rise to $\mathcal{O}_{8}$ and $\mathcal{O}_{9}$, suggesting that these operators should be considered together when placing constraints.

\section{Details of loop-level anapole dark matter models} \label{app:anapoleformfactor}

\subsection{Anapole Moments}
In this subsection, we define the anapole moment of $\chi$ generated at loop-level by a scalar ($\phi$) or vector ($V$) mediator accompanied by a fermion ($\psi$). The anapole moment is the low momentum limit of the anapole form factor, which we will explore in the next subsection. The scalar-mediated anapole moment for the process described by the Lagrangian in Equation~\ref{eqn:minimal_anapole}, is~\cite{Ibarra:2022nzm}
\begin{equation} \label{eqn:anapole_moment}
    \mathcal A_S = \frac{-e Q}{96\pi^2 m^2_\chi}\big[|y_L|^2 - |y_R|^2 \big]\mathcal F_S(\frac{m_\psi}{m_\chi}, \frac{m_\phi}{m_\chi}).
\end{equation}
Similarly, the vector mediated anapole moment is,
\begin{equation}
    \mathcal A_V = \frac{-e Q}{96\pi^2 m^2_\chi}\bigg[2\big[|y_L|^2 - |y_R|^2 \big]\mathcal F_V(\frac{m_\psi}{m_\chi}, \frac{m_\phi}{m_\chi}) +\big[|y^G_L|^2 - |y^G_R|^2 \big]\mathcal F_S(\frac{m_\psi}{m_\chi}, \frac{m_\phi}{m_\chi})\bigg] . 
\end{equation}
The auxiliary functions $\mathcal{F}_i$ are defined as
\begin{equation}
    \mathcal{F}_i(\mu, \eta) = \frac{3}{2}\log\frac{\mu^2}{\eta^2} + (3\eta^2 - 3\mu^2 + n_i)f(\mu, \eta), 
\end{equation}
with $i=S,V,$ and $n_V=-7$, $n_S=1$, and the function $f$ given by
\begin{equation}
    f(\mu, \eta) = \begin{cases} \frac{1}{2\sqrt{\Delta}}\log\frac{\mu^2+\eta^2-1+\sqrt{\Delta}}{\mu^2+\eta^2-1-\sqrt{\Delta}} & \Delta \neq 0 \\ \frac{2}{(\mu^2-\eta^2)^2-1} & \Delta=0\end{cases}
\end{equation}
with $\Delta =(\mu^2-\eta^2-1)^2-4\mu^2\eta^2$.

\subsection{Momentum Dependence of anapole form factor}
\begin{figure*}
    \centering
    \includegraphics[width=0.5\linewidth]{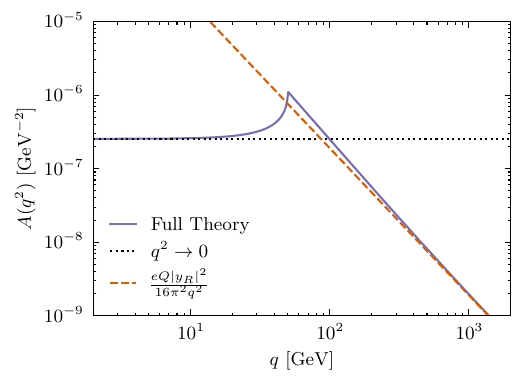}
    \caption{Anapole form factor for $m_\chi = 1~\textrm{GeV}$,  $m_\phi=m_\psi=25~\textrm{GeV}$, and $Q=1, y_L=0, y_R=1$. The solid line shows the full momentum dependence, computed using Equation~\ref{eqn:anapoleFF}. The dotted line shows the anapole moment $\mathcal{A}_S$ using~\ref{eqn:anapole_moment} for the same parameters. The dashed line is the expression $\frac{eQ|y_R|^2}{16\pi^2q^2}$, the asymptotic value of A($q^2$) when $q^2\gg m^2_\chi, m^2_\phi, m^2_\psi$.}
    \label{fig:anapoleFF}
\end{figure*}

In~\cite{Sandick:2016zut}, the authors calculated an analytic expression for the anapole form factor retaining the full momentum and mass dependence in the amplitudes. The scalar-mediated anapole form factor can be expressed in terms of the 3-point functions $C_{(..)}$ as defined in the \textsc{LoopTools}~\cite{Hahn:1998yk} package as, 
\begin{align}\label{eqn:anapoleFF}
    A_S(q^2)= \frac{-e Q}{16\pi^2 m^2_\chi}\big[|y_L|^2 - |y_R|^2 \big]\times\bigg[&C_{11}(m^2_\chi, q^2, m^2_\chi, m^2_\phi, m^2_\psi, m^2_\psi) \\ \nonumber&+ C_{11}(m^2_\chi, q^2, m^2_\chi, m^2_\psi, m^2_\phi,  m^2_\phi) +  C_{1}(m^2_\chi, q^2, m^2_\chi, m^2_\phi, m^2_\psi, m^2_\psi) \\ \nonumber &- C_{12}(m^2_\chi, q^2, m^2_\chi, m^2_\phi, m^2_\psi, m^2_\psi) - C_{12}(m^2_\chi, q^2, m^2_\chi, m^2_\psi, m^2_\phi, m^2_\phi)\bigg] .
\end{align}
When $q^2 \gg m^2_\chi,m^2_\psi, m^2_\phi$, this asymptotes to:
\begin{equation} \label{eqn:high_momentum_limit_anapoleFF}
    A_S(q^2) \underset{q^2\rightarrow\infty}{\rightarrow} \frac{eQ(|y_L|^2 - |y_R|^2)}{16\pi^2q^2}
\end{equation}
In Figure~\ref{fig:anapoleFF}, we compare the full anapole form factor from Equation~\ref{eqn:anapoleFF} to the low-momentum limit from Equation~\ref{eqn:anapole_moment}, and the high-momentum limit from~\ref{eqn:high_momentum_limit_anapoleFF}. The same analysis can be performed for the vector-mediated anapole form factor. 

\subsection{$e^+e^- \rightarrow\chi \chi$ cross-section}
The charged states introduced in Equation~\ref{eqn:minimal_anapole} will generically couple to the Z boson with a similar anapole-like interaction. In particular, the new charged states are singlets under $SU(2)_{\rm L}$ and therefore the form factor for the anapole-like interaction is related to the anapole form-factor by $A_Z(q^2) = A(q^2)\tan \theta_w$, where $\theta_w$ is the Weinberg angle.
The production of $\chi\chi$ pairs can proceed through an intermediate photon or Z boson, and be expressed as a sum of the cross sections of these two channels plus an interference term, where each term is defined as, 
\begin{align} \label{eqn:prodcrosssection}
\sigma^\gamma_{e^+e^- \rightarrow \chi\chi}&=\frac{\alpha}{3s}A(s) s^2 \bigg(1-\frac{4m^2_\chi}{s} \bigg)^{3/2} \\ \nonumber\sigma^Z_{e^+e^- \rightarrow \chi\chi} &= \frac{1}{\cos^4\theta_w}\frac{\alpha A(s) s^2}{12}\frac{s(v^2_e +a^2_e)}{(s-M^2_Z)^2+M^2_Z\Gamma^2_Z} \bigg(1-\frac{4m^2_\chi}{s}\bigg)^{3/2} \\ \nonumber
\sigma^{\rm int}_{e^+e^- \rightarrow \chi\chi} &= \frac{1}{\cos^2\theta_w}\frac{\alpha v_e A(s) s^2 }{3}\frac{(s-M^2_Z)}{(s-M^2_Z)^2+M^2_Z\Gamma^2_Z}\bigg(1-\frac{4m^2_\chi}{s}\bigg)^{3/2} ,
\end{align}
where $v_e$ and $a_e$ are the vector and axial couplings of the electron to the Z boson, respectively. Note that we have assumed the electron to be massless in these expressions.

\begin{figure*} 
    \centering
    \includegraphics[width=0.5\linewidth]{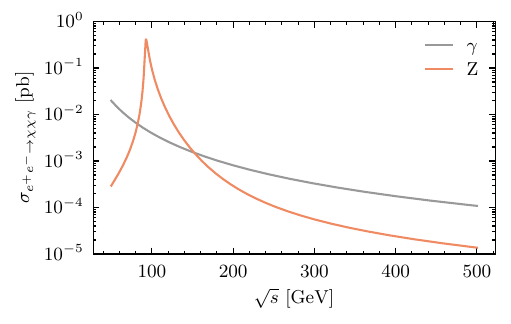}
    \caption{Production cross section of $\chi\chi$ pairs + an ISR photon in the kinematic region selected in~\cite{L3:2003yon} as a function of $\sqrt{s}$, assuming $m_\chi=1~\mathrm{GeV}$, $m_{\mathrm{heavy}}=20~\mathrm{GeV}$, $Q=1$ and $y = \sqrt{4\pi}$. }
    \label{fig:anapole_xsec_s}
\end{figure*}

Pair produced $\chi$ are an invisible final state at colliders, so the relevant experimental signature is an ISR photon plus missing transverse energy (MET). At the energies we are interested in, the differential cross section $\sigma_{e^=e^-\rightarrow \chi \chi \gamma}$ approximately factorizes into the tree level cross section multiplied by the probability $R(x_\gamma, \theta_\gamma)$ to emit an ISR photon with $x_\gamma=\frac{E_\gamma}{\sqrt{s}}$ at an angle $\theta_\gamma$ in the CM frame~\cite{Nicrosini:1989pn,Montagna:1995wp},
\begin{equation} \label{eq:chichigammaxsec}
    \frac{d^2\sigma}{dx_\gamma d\cos\theta} = \sigma_{e^+e^-\rightarrow \chi \chi}(s_{\chi\chi})\times R(x_\gamma, \theta_\gamma, s), 
\end{equation}
where $s_{\chi\chi} = (1-x_\gamma)s$, and
\begin{equation}
    R(x_\gamma, \theta_\gamma, s) = \frac{\alpha}{\pi}\frac{1}{x_\gamma}\bigg[\frac{1+(1-x_\gamma)^2}{1+\frac{4m^2_e}{s}-\cos^2\theta_\gamma} -\frac{x^2_\gamma}{2}\bigg] . 
\end{equation}
The expected number of events from a high energy single photon data sets is given by integrating equation~\ref{eq:chichigammaxsec}, in~ Ref. \cite{L3:2003yon} the kinematic region is defined as $x^{\rm min}_\gamma = 0.02$, $x^{\rm max}_\gamma = 1.0$ and $\theta^{\rm min}_\gamma = 14~\deg$, $\theta^{\rm max}_\gamma = 166~\deg$. We show in Figure~\ref{fig:chichigammaxsec} the integrated cross section as a function of $M$ in this kinematic region at $\sqrt{s} = 200~\mathrm{GeV}$ including the contributions from the Z boson. Figure~\ref{fig:anapole_xsec_s} shows the relative contributions to the total cross as a function of energy, assuming $m_\chi = 1~\rm{GeV}$, $m_{\mathrm{heavy}}=20~\rm{GeV}$, $y=\sqrt{4\pi}$, and $Q=1$. As expected, there is a large enhancement in the vicinity of $M_Z\approx91~\rm{GeV}$. By $\sqrt{s}\approx 200~\rm{GeV}$, the Z boson channel only makes up $\sim20\%$ of the total cross section.

\bibliographystyle{apsrev4-1}
\bibliography{biblio}

\end{document}